\documentclass[%
 reprint,
superscriptaddress,
 amsmath,amssymb,
 aps,
 prl,
]{revtex4-1}

\usepackage{graphicx}% Include figure files
\usepackage{dcolumn}% Align table columns on decimal point
\usepackage{bm}% bold math
\begin{document}

\preprint{APS/123-QED}

\title{Optical Signatures of Quantum Emitters in Suspended Hexagonal Boron Nitride}% Force line breaks with \\

\author{Annemarie L. Exarhos}
\affiliation{Quantum Engineering Laboratory, Department of Electrical and Systems Engineering, University of Pennsylvania, Philadelphia, Pennsylvania 19104, United States}

\author{David A. Hopper}
\affiliation{Quantum Engineering Laboratory, Department of Electrical and Systems Engineering, University of Pennsylvania, Philadelphia, Pennsylvania 19104, United States}
\affiliation{Department of Physics, University of Pennsylvania, Philadelphia, Pennsylvania 19104, United States}

\author{Richard R. Grote}
\affiliation{Quantum Engineering Laboratory, Department of Electrical and Systems Engineering, University of Pennsylvania, Philadelphia, Pennsylvania 19104, United States}

\author{Audrius Alkauskas}
\affiliation{Center for Physical Sciences and Technology, Vilnius LT-01108, Lithuania}
\affiliation{Department of Physics, Kaunas University of Technology, Kaunas LT-51368, Lithuania}

\author{Lee C. Bassett}
\email{lbassett@seas.upenn.edu}
\affiliation{Quantum Engineering Laboratory, Department of Electrical and Systems Engineering, University of Pennsylvania, Philadelphia, Pennsylvania 19104, United States}

\begin{abstract}
Hexagonal boron nitride (h-BN) is a tantalizing material for solid-state quantum engineering.  Analogously to three-dimensional wide-bandgap semiconductors like diamond, h-BN hosts isolated defects exhibiting visible fluorescence, and the ability to position such quantum emitters within a two-dimensional material promises breakthrough advances in quantum sensing, photonics, and other quantum technologies. Critical to such applications, however, is an understanding of the physics underlying h-BN's quantum emission.  We report the creation and characterization of visible single-photon sources in suspended, single-crystal, h-BN films.  The emitters are bright and stable over timescales of several months in ambient conditions.  With substrate interactions eliminated, we study the spectral, temporal, and spatial characteristics of the defects' optical emission, which offer several clues about their electronic and chemical structure.  Analysis of the defects' spectra reveals similarities in vibronic coupling despite widely-varying fluorescence wavelengths, and a statistical analysis of their polarized emission patterns indicates a correlation between the optical dipole orientations of some defects and the primitive crystallographic axes of the single-crystal h-BN film. These measurements constrain possible defect models, and, moreover, suggest that several classes of emitters can exist simultaneously in free-standing h-BN, whether they be different defects, different charge states of the same defect, or the result of strong local perturbations. 
\end{abstract}

\maketitle

%____________________________________________________
%	MANUSCRIPT
%____________________________________________________
%define the manuscript figures

\def\PLimages{\begin{figure}[t]
\centering
%\fbox{
 \includegraphics[angle=0,scale=1,trim=0in 0in 0in 0in,clip=true]{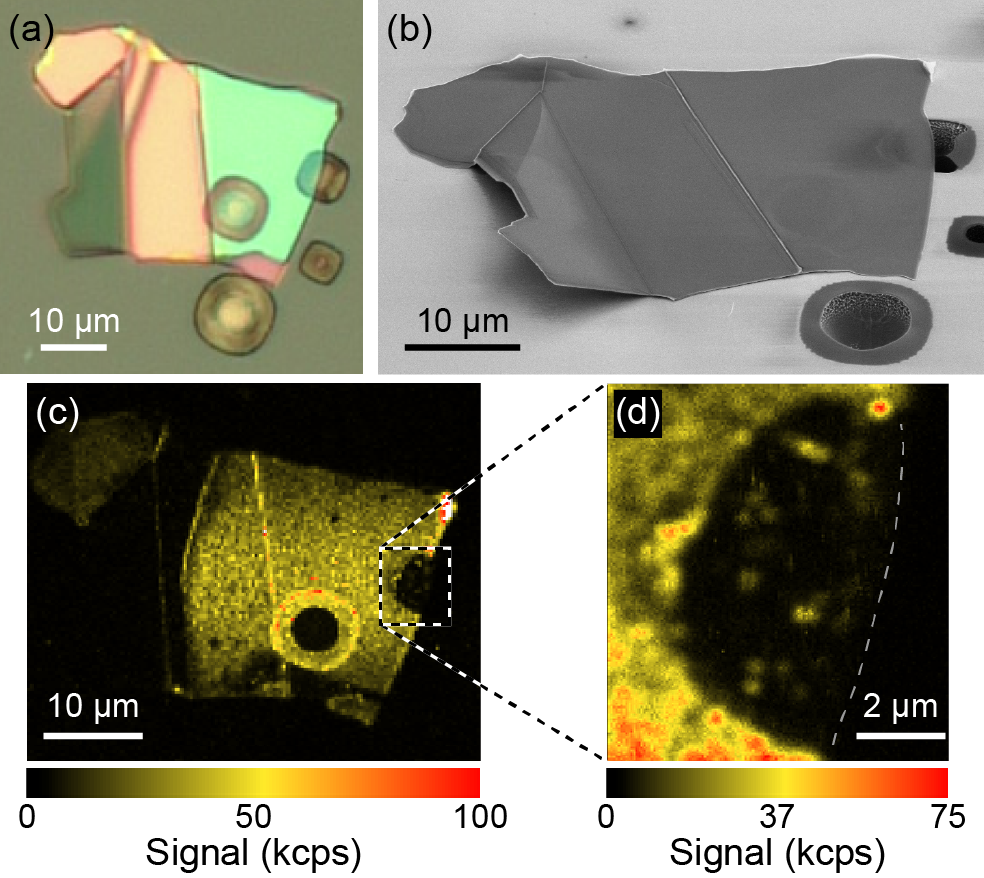}
%} %l, b, r, t
\caption{\textbf{Visible fluorescence from exfoliated h-BN.} \textbf{(a)}  White light optical micrograph of exfoliated h-BN on a patterned Si/SiO$_2$ substrate.  \textbf{(b)}  Tilted SEM image of the same flake.  \textbf{(c)}  PL image of the exfoliated flake under 532 nm excitation.   \textbf{(d)}  PL image of the partially suspended film denoted by the square in (c).  The dashed line shows the edge of the suspended h-BN flake which hangs over the center of the etched hole.}
  \label{PLimages}
\end{figure}}

\def\singleemitters{\begin{figure*}[t]
\centering
%\fbox{
 \includegraphics[angle=0,scale=1,trim=0.0in 0in 0in 0in,clip=true]{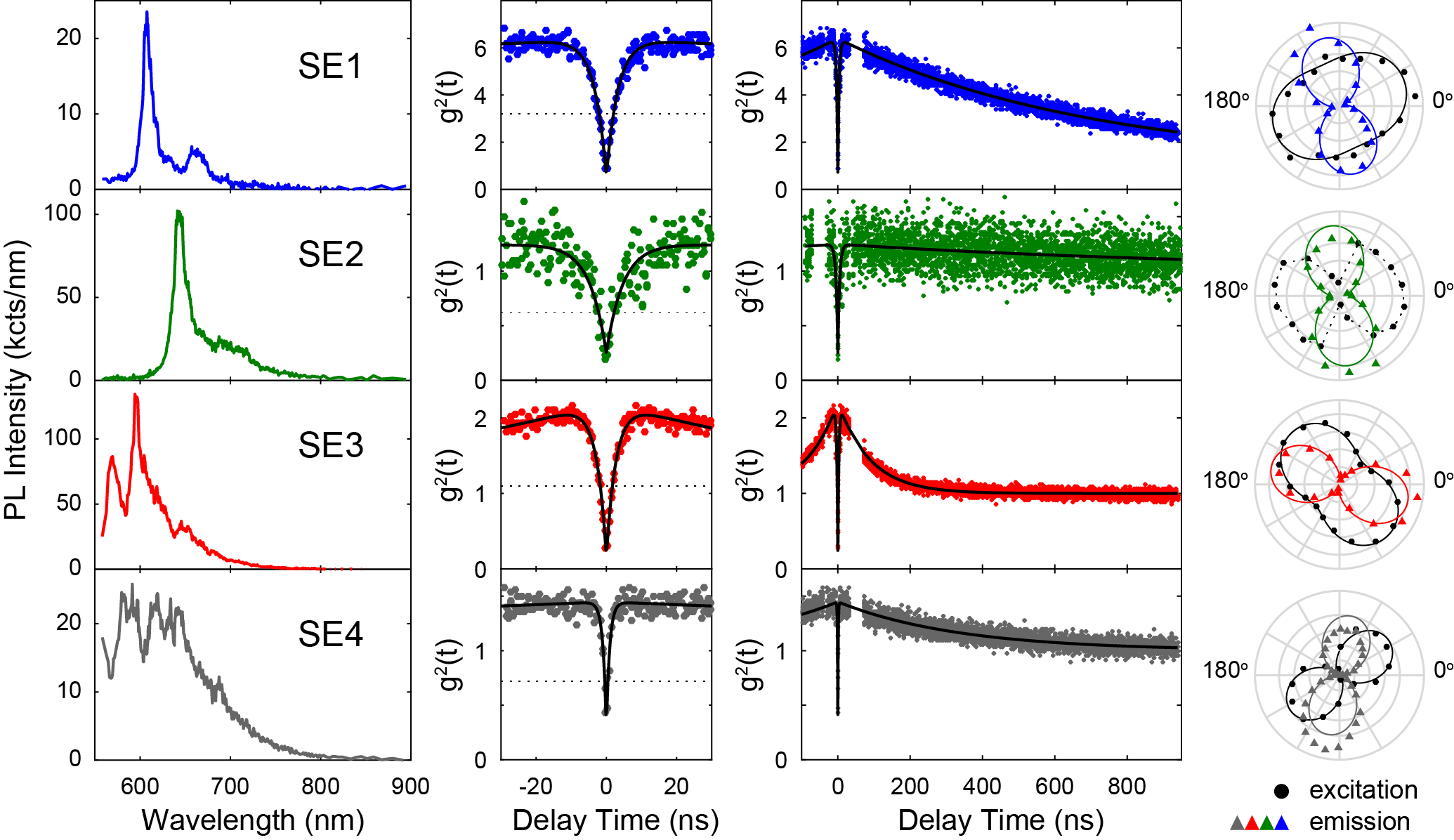}
% }
  \caption{ \textbf{Single emitters in free-standing h-BN.} Panels from left to right: PL spectra, photon autocorrelation measurements over short (30 ns) and long (1~$\mu$s) timescales, and PL excitation and emission polarization dependences.  Each row corresponds to a different single-photon emitter. Dotted lines in the short-time autocorrelation plots indicate the single-emitter threshold. Gaps in the long timescale autocorrelation data correspond to regions where detector crosstalk afterflashes interfere with data collection. }
  \label{singleemitters}
\end{figure*}}

\def\PSBanalysis{\begin{figure*}[t]
\centering
%\fbox{
 \includegraphics[angle=0,scale=1,trim=.07in 0in 0in 0in,clip=true]{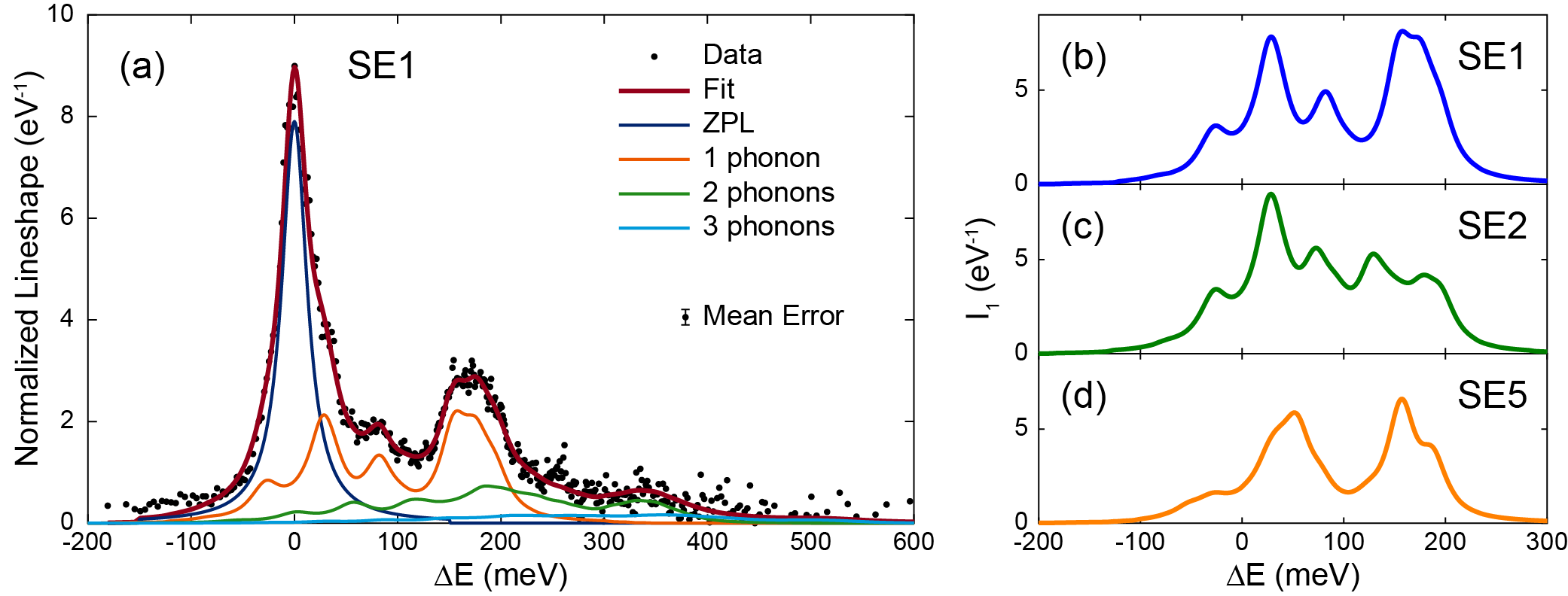}
% }
\caption{\textbf{Vibronic analysis of single-emitter spectra.} \textbf{(a)} Emission lineshape for SE1 (points), as a function of the change in lattice energy during optical relaxation ($\Delta E = E_\mathrm{ZPL}-E$, where $E$ is the observed photon energy). The data are binned to produce approximately uniform uncertainty, as indicated by the representative error bar. Curves show the results of a fit using the model described in the text (thick red curve), along with the ZPL and PSB components (thin curves) as indicated by the legend. \textbf{(b)} Best-fit one-phonon probability distribution functions for emitters SE1, SE2, and SE5. }
  \label{PSBanalysis}
\end{figure*}}

\def\polarization{\begin{figure*}[t]
\centering
%	\fbox{
 \includegraphics[angle=0,scale=1,trim=0.065in 0in 0.095in 0in,clip=true]{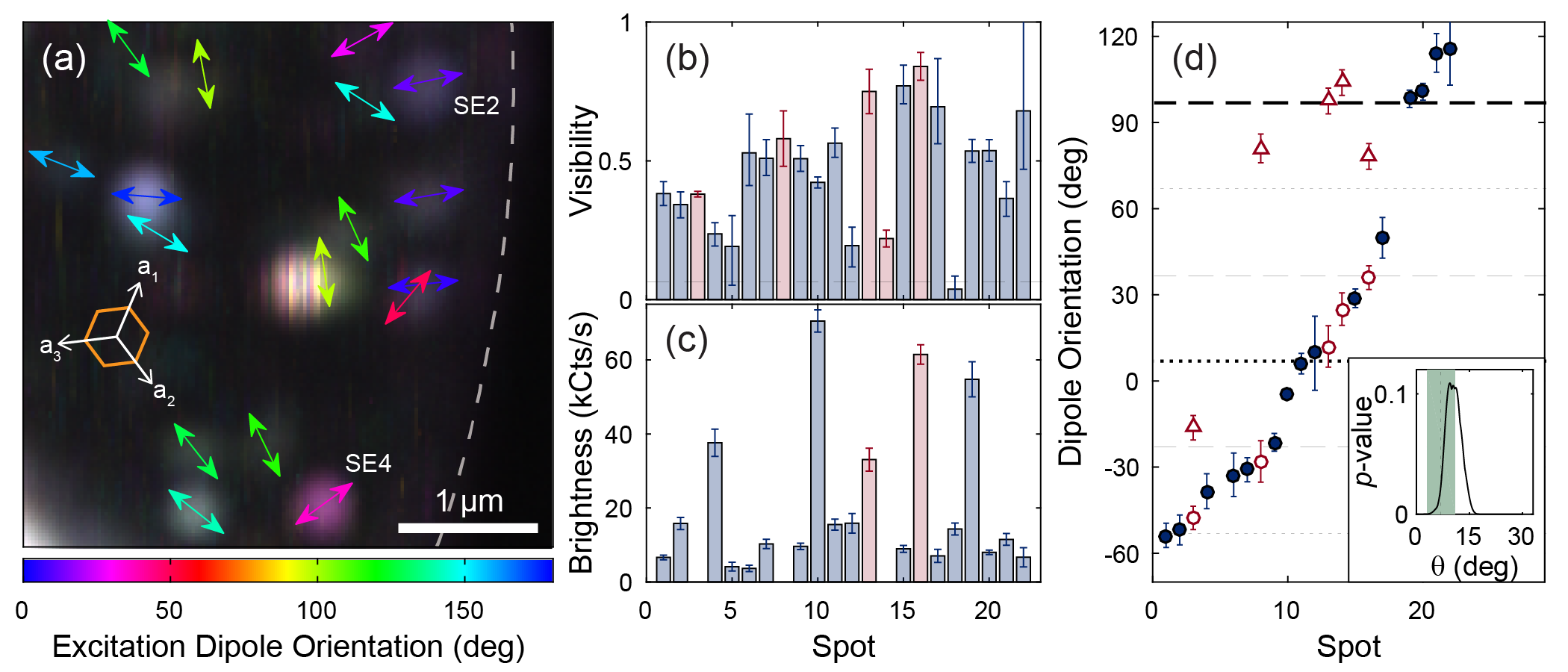}
% }
    \caption{\textbf{Polarization dependence of visible emitters in suspended h-BN.}  \textbf{(a)} Color-coded image indicating the direction (color) and degree (saturation) of PL excitation polarization dependence.  Arrows indicate the orientation of the excitation dipole extracted from fits to the polarization-dependent PL intensity for each spot.  The h-BN crystal orientation is noted. The dotted line corresponds to the edge of the h-BN film over the suspended region.  Emitters without arrows correspond to those for which excitation polarization dependence was unable to be obtained (e.g. due to rapid blinking).  \textbf{(b-d)} Excitation dipole visibility (b), brightness (c), and orientation (d, filled circles) for the spots identified in (a).    Confirmed single-photon sources are colored red, and open circles (triangles) in (d) denote their excitation (emission) dipole orientations.  Separate measurements for SE1, SE3, and SE5 are included in (b) and (d). The dashed line in (b) indicates the PL background visibility. Dotted (dashed) lines in (d) denote angles parallel (perpendicular) to the h-BN $\langle 11\bar{2}0\rangle$ axes. Inset to (d): $p$-value test statistic for a model of dipole alignment to h-BN crystal directions, as a function of the crystal rotation angle, $\theta$.  The dotted line and green shaded region indicate the EBSD measurement and corresponding uncertainty of $\theta$.}
  \label{polarization}
\end{figure*}}
%______________________________

Defect engineering in solid-state materials is a rapidly progressing field with applications in quantum information science \cite{AwschalomBassettDzurakEtAl2013,Heremans2016},  nanophotonics \cite{Gao2015}, and nanoscale sensing in biology and chemistry \cite{Schirhagl2014}.  Inspired by the success of the archetypal nitrogen-vacancy center in diamond \cite{DohertyMansonDelaneyEtAl2013}, recent efforts have uncovered analogous systems in other wide-bandgap semiconductors such as silicon carbide \cite{Weber11052010,KoehlBuckleyHeremansEtAl2011} which offer exciting new opportunities for defect engineering in three-dimensional materials.  However, optically active impurities in low-dimensional materials and thin films can provide unique functionalities due to intrinsic spatial confinement and the ability to create multi-functional layered materials \cite{GeimGrigorieva2013,Wrachtrup2016}.  Within the class of van der Waals materials, hexagonal boron nitride (h-BN) is an ideal candidate for exploring new defect physics due to its large ($\sim$6 eV) bandgap \cite{Cassabois2016} and its unique optical \cite{Dai2015}, electrical \cite{PhysRevB.80.224301,PhysRevB.83.115328}, and vibronic properties \cite{Dai2014} that may influence the underlying physics of its defects.  At present, however, progress is impeded by an incomplete understanding of the electronic and chemical structure of defects responsible for h-BN's visible fluorescence.

Here, we describe the creation and observation of single-photon sources in suspended, single-crystal h-BN membranes.  Marked differences between the photoluminescence (PL) response of regions freely suspended or supported on Si/SiO$_2$ substrates point to strong substrate-dependent effects in the defects' formation and optical properties.  Individual emitters within the same single-crystal flake of suspended h-BN display a range of distinctly different PL spectra, polarization properties, brightness, and photon emission statistics.  Furthermore, the absorptive and emissive dipoles show a weak statistical alignment with h-BN's crystallographic basis vectors. Together, these measurements support the presence of multiple defect species within h-BN, clarifying conflicting earlier reports and providing a framework for further theoretical and experimental investigations of their properties.

Due to h-BN's crucial role as a two-dimensional (2D) dielectric constituent in layered materials, its defects have received a surge of attention in recent years.  Scanning-tunneling microscopy of graphene/h-BN heterostructures revealed charged impurities attributed to native defects in h-BN possibly involving carbon \cite{Wong2015}. Carbon impurities have also been implicated in ultraviolet photoluminescence \cite{PhysRevB.78.155204} and cathodoluminescence \cite{Meuret2015} from h-BN, even at the single-defect level \cite{Bourrellier2016}.  Recently, single-photon sources at visible wavelengths have been reported from supported monolayer, multilayer, and bulk h-BN \cite{TranBrayFordEtAl2016,PhysRevApplied.5.034005,2016arXiv160609364S,2016arXiv160604124M,2016arXiv160504445J}, but existing observations and interpretations vary widely in terms of basic emitter properties (\textit{e.g.,} optical lifetime, spectral line shape, brightness) and proposed physical models.  By focusing on regions of suspended h-BN, we eliminate substrate interactions that are observed to play an important role in the material's visible emission characteristics.  Additionally, our characterization of multiple defects within a single-crystal film offers new insight into the electronic properties and likely chemical structures of this new class of quantum emitters.

Single-crystal h-BN is mechanically exfoliated \cite{PacileMeyerGiritEtAl2008} onto patterned Si substrates capped with a 90 nm layer of thermal oxide chosen for optimal contrast in bright-field optical microscopy \cite{SMLL:SMLL201001628}.  Holes are created in the support wafer by optical lithography followed by dry etching.  The process results in three distinct regions of the patterned substrate: Si/SiO$_2$, free-standing SiO$_2$, and etched holes several microns wide and $\sim$5 $\mu$m deep.  Exfoliated h-BN flakes are typically $\sim$1000 $\mu$m$^2$ in area and extend over all three regions.  We confirm that the exfoliated h-BN maintains the single-crystallinity of its parent crystal \cite{Novoselov26072005} using EBSD, which additionally determines the in-plane crystallographic orientation of exfoliated flakes.  Film thickness is determined using atomic force microscopy; the flake studied here is 150~nm thick in the vicinity of the suspended region.  Raman spectroscopy confirms that this flake exhibits bulk behavior, as expected for a membrane containing several hundred atomic layers \cite{PhysRevB.71.205201}.  Exfoliated samples are annealed in an argon atmosphere, followed by an O$_2$ plasma treatment, electron bombardment \emph{via} a scanning electron microscope, and an additional argon anneal.  A home-built confocal scanning fluorescence microscope is used to acquire PL images, spectra, and photon emission statistics, where the collected PL is directed either to a pair of silicon single-photon avalanche diodes or a spectrometer.  Additional details can be found in the Supplementary Materials \cite{Supplement}.

\PLimages
Fig.~\ref{PLimages} shows an exfoliated, single-crystal h-BN flake under various imaging modalities.  Bright-field optical microscopy (Fig.~\ref{PLimages}a) shows the holes etched into the substrate beneath the flake, and scanning electron microscope (SEM) images  (Fig.~\ref{PLimages}b) confirm that the flake suspends the holes.  Spatial PL images (Fig.~\ref{PLimages}c-d) taken with 532~nm excitation exhibit strong PL from the supported h-BN as compared to both the suspended regions and bare Si/SiO$_2$ substrate.  A higher-resolution image (Fig.~\ref{PLimages}d) of the partially covered hole denoted in Fig.~\ref{PLimages}c reveals bright spots within the free-standing h-BN membrane that we characterize as single- and few-photon emitters.  No isolated single emitters were observed in the suspended regions prior to electron beam exposure and successive Ar anneal \cite{Supplement}, although the supported regions of the flake always displayed some PL.  Other suspended regions on the same flake do not show evidence of defect PL despite undergoing the same treatment.  
  
We find a significant difference between the brightness of supported versus suspended regions, where the supported region is typically $\sim$5--10 times brighter than the suspended region.  This difference in brightness appears to imply substrate-dependent defect formation, with a higher defect density observed on supported regions, but the apparent brightness of individual emitters seems to depend on the substrate as well.  The brightest emitters in the suspended region approach the average brightness of those on the supported region, while the vast majority of suspended emitters are dimmer than their supported counterparts (see Figs.~S6-S7 \cite{Supplement}).  Additionally, the PL intensity from supported h-BN varies as a function of the substrate conditions, with the brightest regions occurring where h-BN overlays a released SiO$_2$ membrane at the edge of a hole, rather than the full Si/SiO$_2$ heterostructure.  These puzzling effects are visible in Fig.~\ref{PLimages}c and much more apparent in Fig.~S6.  Optical interference effects might play a role, but substrate-induced brightening of individual emitters is still surprising, since substrate interactions are generally supposed to quench rather than enhance defect PL.

\singleemitters
The PL characteristics for four single-photon sources in the freely suspended h-BN membrane are summarized in Fig.~\ref{singleemitters}.  Single-photon emission is established by the presence of an antibunching dip near zero delay in the photon autocorrelation function, $g^{(2)}(t)$, that drops below the threshold indicated by a dotted line in each panel of the second column in Fig.~\ref{singleemitters} \cite{Supplement}. The room-temperature emission spectra of single defects (leftmost column of Fig.~\ref{singleemitters}) exhibit striking differences both in shape and spectral weight, with features extending between 550-700 nm (1.77-2.25 eV).  Some spectra, such as for SE1 and SE2, include a clear zero-phonon line (ZPL) and phonon sideband (PSB) and are similar in shape to those reported from experiments with supported h-BN samples \cite{TranBrayFordEtAl2016,PhysRevApplied.5.034005,2016arXiv160609364S,2016arXiv160604124M,2016arXiv160504445J}.  Other spectra, such as for SE3 and SE4, display spectral features distinct from what has been thus far reported.  The emission of SE4, in particular, is characterized by a much broader spectral distribution than for SE1 and SE2, with a comparatively high-energy ZPL that appears to be outside our detection bandwidth.  Measurements of a fifth single emitter (SE5) in the suspended h-BN region with broadly similar characteristics to SE1 and SE2 are provided in the Supplementary Materials \cite{Supplement}.

\PSBanalysis
For the emitters whose PL spectra exhibit a clearly discernible ZPL (SE1, SE2, and SE5), we model the emission spectra using the well-established theory for electron-phonon coupling of defects in solids \cite{Maradudin1966,Davies1974,Supplement}.   Fig.~\ref{PSBanalysis}a presents a fit of this model to the emission lineshape of SE1, which is derived from the corresponding spectrum in Fig.~\ref{singleemitters} by the spectral density to energy units and accounting for the $E^3$ dependence of spontaneous emission, where $E$ is the photon energy.  From the fit we extract the ZPL energy, $E_\mathrm{ZPL}$, the full width at half maximum of the Lorentzian ZPL lineshape, $\Gamma_\mathrm{ZPL}$, and the Huang-Rhys factor, $S_\mathrm{HR}$, for each defect.  The best-fit values for these parameters are presented along with other data in Table~\ref{table}. The fit also returns the functional form of the $n$-phonon contributions to the PSB, $I_n(\Delta E)$.  For example, $I_1(\Delta E)$ is the probability distribution function describing the change in lattice vibrational energy, $\Delta E$, due to the emission ($\Delta E>0$) or absorption ($\Delta E<0$) of one phonon during optical relaxation.  The best-fit distributions $I_1(\Delta E)$ for SE1, SE2, and SE5 are plotted in Figs.~3b-d, respectively.  

While we caution that the theory and fitting procedure relies on several approximations, we can draw useful information from qualitative comparisons of the results.  In particular, $S_\mathrm{HR}$ quantifies the strength of electron-phonon interactions for a given defect and $I_1(E)$ highlights the energies of dominant phonon modes.  Interestingly, despite ZPL energies differing by over 100~meV, SE1 and SE2 exhibit similar Huang-Rhys factors ($S_\mathrm{HR}\sim 1$) and PSB structure, with clear peaks around $\Delta E=30$, 75, and 175~meV, albeit with differences between 100-150~meV.  In contrast, SE5 exhibits a larger $S_\mathrm{HR}$ and qualitatively different PSB.  All emitters have room-temperature ZPL linewidths $\sim 30$ meV, several times larger than comparable observations from diamond nitrogen-vacancy centers \cite{PhysRevLett.103.256404, PhysRevLett.107.146403}.  

In the case of SE3 and SE4, fits are poorly constrained since $E_\mathrm{ZPL}$ cannot be identified unambiguously, although the broad spectrum of SE4 suggests a much larger value for $S_\mathrm{HR}$. SE3 was observed to blink during measurements, and we believe the two peaks visible in its spectrum around 568~nm and 595~nm in Fig.~\ref{singleemitters} may represent the distinct ZPLs of two configurations of the same defect.  If that is the case, the individual spectral configurations seem broadly similar to SE1 and SE2. Further details of the spectral analysis can be found in the Supplementary Materials \cite{Supplement}. 

A defect's photon emission statistics can also provide insight into the number and relaxation lifetimes of individual electronic levels involved in its optical cycle.  The central columns of Fig.~\ref{singleemitters} show $g^{(2)}(\tau)$ for each emitter over both short ($|\tau|<30$ ns) and long ($|\tau|<1$ $\mu$s) timescales.  Significant bunching ($g^{(2)}(t)>1$) is observed for some emitters at intermediate delay times, indicating the participation of at least three electronic levels with different lifetimes \cite{PhysRevA.58.620}.  We adopt a fitting function that accounts for three levels as well as experimental nonidealities that lift the observed $g^2(0)$ slightly above zero: 
\begin{equation}
g^2(t) = 1-\rho^2\big[(1+a)e^{-|t|/\tau_1}-ae^{-|t|/\tau_2}\big],
\end{equation} 
where $t$ is the delay time, $a$ is the photon bunching amplitude, $\tau_1$ and $\tau_2$ correspond to the lifetimes of excited states within the electronic structure of the defect at the effective pump rate determined by the excitation laser, and $\rho<1$ accounts for a Poissonian background \cite{Brouri2000,Supplement}.  The  criterion to identify single emitters is $g^2(0) < \frac{1}{2}(1+\rho^2 a)$, as denoted by the dotted line in the second column of Fig.~\ref{singleemitters}.  That criterion is independent of any background, and is satisfied by at least 6 standard deviations for SE1--SE5.  Best-fit values for $\tau_1$ and $\tau_2$ are given in Table~\ref{table}. The short lifetimes ($\tau_1$) vary from $\sim 1$~ns (SE4) to several nanoseconds (SE1 and SE2) and appear to roughly correlate with differences in the spectral shapes.  The variation of long lifetimes ($\tau_2$) is probably influenced by a dependence on the excitation rate, but the obvious presence of metastable states with lifetimes approaching 1~$\mu$s indicates a potential role for spin physics and intersystem crossings in the defects' optical dynamics.

\begin{table*} 
\centering
\caption{\textbf{Properties of single emitters in suspended h-BN.}}
\label{table}
\begin{tabular}{cccccccc}
Emitter & $E_\mathrm{ZPL}$ (eV) & $S_{HR}$ & $\Gamma_{ZPL}$ (meV) & $\tau_1$ (ns) & $\tau_2$ (ns) & $\Delta\theta$ (deg) & $V_{exc}$\\ \hline 
 \noalign{\vskip 3pt} 
SE1 & 2.0405$\pm$0.0003	& 1.0$\pm0.1$	& 30$\pm$1 & 3.33$\pm$0.05	& 710$\pm$10 & 79$\pm$4 & 0.22$\pm$0.03\\
\noalign{\vskip 3pt} 
SE2	&	1.9269$\pm$0.0003 & 1.2$\pm0.1$ & 31$\pm$2 & 4.7$\pm$0.4 & 1100$\pm$700 & 86$\pm$6$^{\ddagger}$ & 0.76$\pm$0.08$^{\S}$\\
\noalign{\vskip 3pt} 
SE3	& - & -	& - & 2.72$\pm$0.04	& 89.1$\pm$0.6 & 31$\pm$1 & 0.38$\pm$0.01\\ 
\noalign{\vskip 3pt} 
SE4 & - & - & - & 0.94$\pm$0.04$^{\dagger}$ & 335$\pm$8 & 42$\pm$2 & 0.84$\pm$0.05\\
\noalign{\vskip 3pt} 
SE5$^\ast$	& 2.0812$\pm$0.0003 & 1.7$\pm$0.1 & 31$\pm$1 & 1.7$\pm$0.2 & - & 71$\pm$10$^{\ddagger}$ & 0.6$\pm$0.1$^{\S}$ \\ \hline 
\multicolumn{6}{l}{$^\ast$\footnotesize{Data presented in the Supplementary Materials.} }\\
\multicolumn{6}{l}{$^{\dagger}$\footnotesize{Lifetime likely limited by detector timing jitter.}}\\
\multicolumn{6}{l}{$^{\ddagger}$\footnotesize{$\theta_{exc}$ determined using the minimum measured intensity ($\theta_{exc}=\theta_{min}+90^\circ$).}}\\
\multicolumn{6}{l}{$^{\S}$\footnotesize{$V_{exc}$ determined using the maximum and minimum measured intensities.}}\\
\end{tabular}
\end{table*}

All five single emitters display polarization dependence for both the excitation absorption (black points) and emission (colored triangles), as shown in the rightmost column of Fig.~\ref{singleemitters}.  Solid curves indicate normalized fits using the function $I_\mathrm{dip}(\theta)=b+A\cos^2\theta$, where $\theta$ corresponds to either the excitation or emission polarization angle, $b$ is the offset, and $A$ is the amplitude. For SE2, the excitation polarization dependence differs dramatically from the usual emission pattern expected for a single dipole.  Instead, it  exhibits relatively little polarization contrast except in a narrow range of angles where the emission is essentially extinguished.  The dotted line for the excitation polarization dependence of SE2 in Fig.~\ref{singleemitters} is not a fit.  Rather, it serves as a guide to the eye emphasizing this unusual behavior.  SE5 exhibits similar features \cite{Supplement}.

The angle between the excitation and emission dipoles, $\Delta\theta$, as determined by fits to $I_\mathrm{dip}(\theta)$, is listed in Table~\ref{table} along with the excitation visibility, $V_{exc}=A/(2b+A)$.  While the excitation visibilities range widely from 20\% to over 80\%, the polarization visibility in emission appears to be ideal for all five single emitters, within the resolution and bandwidth of our measurements \cite{Supplement}.  Unity visibility suggests that emission occurs \textit{via} a single in-plane electric dipole transition.  However, all emitters exhibit severely misaligned excitation and emission angles, with $\Delta\theta$ ranging from 30$^\circ$ to nearly $90^\circ$. 

The variations in excitation visibility and strong misalignment between absorptive and emissive dipoles could have multiple explanations.  Reorientation between different atomic configurations is possible, particularly for flexible molecular structures with multiple conformations, although it seems less likely for point defects composed of a few atoms in crystalline host to exhibit large dipole orientation shifts due to atomic reconfiguration.  On the other hand, if a dipole possesses a significant out-of-plane component, relatively small deviations could produce large apparent angular shifts in the in-plane projections we measure in this study.  The presence of such canted dipoles might also explain the large variation in brightness we observe between emitters.  If any such atomic reconfigurations occur, however, they must be highly reproducible in order to explain the stable, high-visibility, polarized emission from all defects we have observed.

Alternatively, the presence of multiple excited states with different symmetries can mean that absorption probes a transition associated with the higher-lying state, while emission occurs only from the lower-energy state at a different polarization \cite{Lakowicz}.  This occurs, for example, in the case of the diamond silicon-vacancy center \cite{PhysRevB.89.235101}. Multiple electronic levels are also indicated by the bunching observed in photon autocorrelation measurements, discussed above, suggesting these defects exhibit a relatively complex level structure. For all confirmed single emitters, we observe that the full defect spectrum varies uniformly with excitation polarization.  However, since the emission polarization dependence is measured only for a relatively narrow PL band ($\lambda_{PL} \in[600 \mbox{ nm, } 650 \mbox{ nm}]$), it may not capture energy-dependent variations in the PL polarization. Future photoluminescence excitation (PLE) measurements as a function of excitation and emission energy are needed to clarify this issue. 

\polarization
While only five spots in the present suspended film were sufficiently isolated and bright to confirm as single emitters \textit{via} autocorrelation measurements, all spots for which $g^{(2)}(t)$ was recorded show evidence of antibunching.  The polarization and spectral signatures of other features in Fig.~\ref{PLimages}d also suggest that most spots are single emitters.  Fig.~\ref{polarization}a is a composite image created from multiple spatial PL maps of a region of the suspended flake, recorded at various linear excitation polarizations between $0^{\circ}$ and $180^{\circ}$.  The individual images are color filtered and summed such that the resulting color, value, and saturation correspond to excitation dipole orientation, visibility, and PL brightness, respectively \cite{PhysRevB.76.165205}.
Such images are helpful in identifying candidate single emitters based on their visibility and brightness.  

To quantify the polarization measurements in Fig.~\ref{polarization}a, we fit each PL image to a set of 2D Gaussian peaks to extract the amplitude of each identifiable spot, and then fit each spot's intensity variation to the dipole formula $I_\mathrm{dip}(\theta)$.  The best-fit visibility, peak brightness, and excitation dipole angle for 19 isolated spots in Fig.~\ref{polarization}a are plotted in Fig.~\ref{polarization}b-d, and the excitation dipole angles are superimposed as arrows in Fig.~\ref{polarization}a.  SE1, SE3, and SE5 do not appear in Fig.~\ref{polarization}a, but separate measurements of visibility and dipole orientation are included in Fig.~\ref{polarization}b and~4d (the brightness is not comparable since these measurements were acquired under different illumination conditions).  Figs.~4b-c highlight the significant variation in both the visibility and brightness of emitters in free-standing h-BN.  Notably, this variation extends to the confirmed single-photon emitters (red bars) as well. 

All emitters in Fig.~\ref{polarization} originate from the same suspended flake of single-crystal h-BN, which presents an opportunity to look for correlations between the observed dipole orientations and the host's crystallographic axes.  Such alignments are generally expected for simple native defects (e.g., vacancies, substitutional atoms, and their binary complexes) based on symmetry considerations.  The in-plane orientation of the flake's crystallographic axes, determined using electron backscatter diffraction (EBSD) \cite{Supplement}, are plotted in Fig.~\ref{polarization}a. The vectors $\{\mathbf{a}_1,\mathbf{a}_2,\mathbf{a}_3\}$ are parallel to the equivalent $\langle 11\bar{2}0\rangle$ hexagonal lattice vectors, corresponding to bonds connecting in-plane B and N atoms in h-BN. By symmetry, undistorted mono- or di-atomic defects can give rise to electric dipole matrix elements parallel or perpendicular to these crystallographic axes.  Furthermore, the typical $AA'$ stacking of multilayer h-BN \cite{PhysRevLett.111.036104} entails a 60$^\circ$ rotation between alternate layers.  This means that the allowed orientations for in-plane dipoles of simple defects in single-crystal h-BN should be spaced at 30$^\circ$ increments.  Those possible orientations are indicated in Fig.~\ref{polarization}d by dashed and dotted lines.  

A cursory examination of the data does not reveal any obvious correlation between the observed dipole angles and primitive crystallographic vectors.  Rather, the dipoles appear to be nearly uniformly distributed at random angles.  Nonetheless, statistical analysis does provide tentative evidence for dipole-lattice alignment, at least for a subset of the observations.  The inset to Fig.~\ref{polarization}d plots the $p$-value test statistic corresponding to a lattice-locking model in which each dipole is assumed to align along or orthogonal to the nearest allowed lattice vector (spaced with a 30$^\circ$ period), as a function of the crystal rotation angle, $\theta$. In calculating the $\chi^2$ fit statistic, we retain the excitation and emission dipoles of all confirmed single emitters, but exclude five out of the fifteen observations from unconfirmed spots (chosen independently for each crystal orientation), on the basis that they could be overlapping defects \cite{Supplement}.  The $p$-value calculated from $\chi^2$ is interpreted as the probability that the data could represent a random sample of observations derived from the model.  The fact that the $p$-value exceeds the conventional threshold of 0.05 for some alignment angles indicates that we cannot statistically exclude lattice locking based on these observations.  Moreover, the close agreement between the best-fit angle, $\theta_\mathrm{Fit}=(9.5^{+2.6}_{-1.0})^\circ$, and the EBSD-determined orientation, $\theta_\mathrm{EBSD}=(7\pm4)^\circ$,  suggests that lattice locking might indeed play a role, \textit{i.e.}, the observed alignment of some dipoles to the crystal axes might not be coincidental.

This statistical analysis includes the polarization dependence of all five confirmed single emitters.  However, their disparate spectral and temporal characteristics (Fig.~\ref{singleemitters}) suggest that h-BN hosts single-photon sources with several different defect structures.  The varied dipole orientations and (mis)alignment with the underlying lattice (Fig.~\ref{polarization}d) further supports this idea.  For example, SE4 and SE5 have well-defined emission dipoles that are clearly not aligned with the crystallographic axes, which might indicate structural deformations or more complex chemistries for these defects. Finally, we observe that Fig.~\ref{polarization}a hints at possible correlations between dipole orientation and defect position on the suspended h-BN membrane. Emitters near the edge of the membrane (dashed line in Fig.~\ref{polarization}a) appear to have similar excitation dipole orientations and are generally different from emitters elsewhere in the region.  These effects might be influenced by strain variations across the suspended sample, something that would play an even larger role as the h-BN thickness decreases towards the 2D limit.  Future studies regarding the orientation statistics of similar defects, ideally classified by their spectral and temporal emission properties, will help ascertain the nature of dipole-lattice coupling.

%The observation that substrate interactions can influence both formation and nature of these defects invites further study.
This work indicates that h-BN offers a rich assortment of single-photon sources whose physics can be explored and harnessed.  New insights regarding their optical properties, dipole-lattice coupling, and the importance of substrate interactions resolve some outstanding questions and motivate further study of this fascinating new class of quantum emitters.  As a key building block in functional van der Waals materials, the availability of stable, highly localized, optically addressable electronic states in h-BN offers exciting new opportunities for quantum science and engineering, where the integration of single-photon sources in 2D materials can enable new and unique functionalities.

%\bibliography{Exarhos_bibliography}

\begin{thebibliography}{38}%
\makeatletter
\providecommand \@ifxundefined [1]{%
 \@ifx{#1\undefined}
}%
\providecommand \@ifnum [1]{%
 \ifnum #1\expandafter \@firstoftwo
 \else \expandafter \@secondoftwo
 \fi
}%
\providecommand \@ifx [1]{%
 \ifx #1\expandafter \@firstoftwo
 \else \expandafter \@secondoftwo
 \fi
}%
\providecommand \natexlab [1]{#1}%
\providecommand \enquote  [1]{``#1''}%
\providecommand \bibnamefont  [1]{#1}%
\providecommand \bibfnamefont [1]{#1}%
\providecommand \citenamefont [1]{#1}%
\providecommand \href@noop [0]{\@secondoftwo}%
\providecommand \href [0]{\begingroup \@sanitize@url \@href}%
\providecommand \@href[1]{\@@startlink{#1}\@@href}%
\providecommand \@@href[1]{\endgroup#1\@@endlink}%
\providecommand \@sanitize@url [0]{\catcode `\\12\catcode `\$12\catcode
  `\&12\catcode `\#12\catcode `\^12\catcode `\_12\catcode `\%12\relax}%
\providecommand \@@startlink[1]{}%
\providecommand \@@endlink[0]{}%
\providecommand \url  [0]{\begingroup\@sanitize@url \@url }%
\providecommand \@url [1]{\endgroup\@href {#1}{\urlprefix }}%
\providecommand \urlprefix  [0]{URL }%
\providecommand \Eprint [0]{\href }%
\providecommand \doibase [0]{http://dx.doi.org/}%
\providecommand \selectlanguage [0]{\@gobble}%
\providecommand \bibinfo  [0]{\@secondoftwo}%
\providecommand \bibfield  [0]{\@secondoftwo}%
\providecommand \translation [1]{[#1]}%
\providecommand \BibitemOpen [0]{}%
\providecommand \bibitemStop [0]{}%
\providecommand \bibitemNoStop [0]{.\EOS\space}%
\providecommand \EOS [0]{\spacefactor3000\relax}%
\providecommand \BibitemShut  [1]{\csname bibitem#1\endcsname}%
\let\auto@bib@innerbib\@empty
%</preamble>
\bibitem [{\citenamefont {Awschalom}\ \emph {et~al.}(2013)\citenamefont
  {Awschalom}, \citenamefont {Bassett}, \citenamefont {Dzurak}, \citenamefont
  {Hu},\ and\ \citenamefont {Petta}}]{AwschalomBassettDzurakEtAl2013}%
  \BibitemOpen
  \bibfield  {author} {\bibinfo {author} {\bibfnamefont {D.~D.}\ \bibnamefont
  {Awschalom}}, \bibinfo {author} {\bibfnamefont {L.~C.}\ \bibnamefont
  {Bassett}}, \bibinfo {author} {\bibfnamefont {A.~S.}\ \bibnamefont {Dzurak}},
  \bibinfo {author} {\bibfnamefont {E.~L.}\ \bibnamefont {Hu}}, \ and\ \bibinfo
  {author} {\bibfnamefont {J.~R.}\ \bibnamefont {Petta}},\ }\href {\doibase
  10.1126/science.1231364} {\bibfield  {journal} {\bibinfo  {journal}
  {Science}\ }\textbf {\bibinfo {volume} {339}},\ \bibinfo {pages} {1174}
  (\bibinfo {year} {2013})}\BibitemShut {NoStop}%
\bibitem [{\citenamefont {Heremans}\ \emph {et~al.}(2016)\citenamefont
  {Heremans}, \citenamefont {Yale},\ and\ \citenamefont
  {Awschalom}}]{Heremans2016}%
  \BibitemOpen
  \bibfield  {author} {\bibinfo {author} {\bibfnamefont {F.~J.}\ \bibnamefont
  {Heremans}}, \bibinfo {author} {\bibfnamefont {C.~G.}\ \bibnamefont {Yale}},
  \ and\ \bibinfo {author} {\bibfnamefont {D.~D.}\ \bibnamefont {Awschalom}},\
  }\href {\doibase 10.1109/JPROC.2016.2561274} {\bibfield  {journal} {\bibinfo
  {journal} {Proceedings of the IEEE}\ }\textbf {\bibinfo {volume} {PP}},\
  \bibinfo {pages} {1} (\bibinfo {year} {2016})}\BibitemShut {NoStop}%
\bibitem [{\citenamefont {Gao}\ \emph {et~al.}(2015)\citenamefont {Gao},
  \citenamefont {Imamoglu}, \citenamefont {Bernien},\ and\ \citenamefont
  {Hanson}}]{Gao2015}%
  \BibitemOpen
  \bibfield  {author} {\bibinfo {author} {\bibfnamefont {W.~B.}\ \bibnamefont
  {Gao}}, \bibinfo {author} {\bibfnamefont {A.}~\bibnamefont {Imamoglu}},
  \bibinfo {author} {\bibfnamefont {H.}~\bibnamefont {Bernien}}, \ and\
  \bibinfo {author} {\bibfnamefont {R.}~\bibnamefont {Hanson}},\ }\href
  {http://dx.doi.org/10.1038/nphoton.2015.58} {\bibfield  {journal} {\bibinfo
  {journal} {Nat. Photonics}\ }\textbf {\bibinfo {volume} {9}},\ \bibinfo
  {pages} {363} (\bibinfo {year} {2015})}\BibitemShut {NoStop}%
\bibitem [{\citenamefont {Schirhagl}\ \emph {et~al.}(2014)\citenamefont
  {Schirhagl}, \citenamefont {Chang}, \citenamefont {Loretz},\ and\
  \citenamefont {Degen}}]{Schirhagl2014}%
  \BibitemOpen
  \bibfield  {author} {\bibinfo {author} {\bibfnamefont {R.}~\bibnamefont
  {Schirhagl}}, \bibinfo {author} {\bibfnamefont {K.}~\bibnamefont {Chang}},
  \bibinfo {author} {\bibfnamefont {M.}~\bibnamefont {Loretz}}, \ and\ \bibinfo
  {author} {\bibfnamefont {C.~L.}\ \bibnamefont {Degen}},\ }\href {\doibase
  10.1146/annurev-physchem-040513-103659} {\bibfield  {journal} {\bibinfo
  {journal} {Annual Review of Physical Chemistry}\ }\textbf {\bibinfo {volume}
  {65}},\ \bibinfo {pages} {83} (\bibinfo {year} {2014})}\BibitemShut {NoStop}%
\bibitem [{\citenamefont {Doherty}\ \emph {et~al.}(2013)\citenamefont
  {Doherty}, \citenamefont {Manson}, \citenamefont {Delaney}, \citenamefont
  {Jelezko}, \citenamefont {Wrachtrup},\ and\ \citenamefont
  {Hollenberg}}]{DohertyMansonDelaneyEtAl2013}%
  \BibitemOpen
  \bibfield  {author} {\bibinfo {author} {\bibfnamefont {M.~W.}\ \bibnamefont
  {Doherty}}, \bibinfo {author} {\bibfnamefont {N.~B.}\ \bibnamefont {Manson}},
  \bibinfo {author} {\bibfnamefont {P.}~\bibnamefont {Delaney}}, \bibinfo
  {author} {\bibfnamefont {F.}~\bibnamefont {Jelezko}}, \bibinfo {author}
  {\bibfnamefont {J.}~\bibnamefont {Wrachtrup}}, \ and\ \bibinfo {author}
  {\bibfnamefont {L.~C.}\ \bibnamefont {Hollenberg}},\ }\href {\doibase
  http://dx.doi.org/10.1016/j.physrep.2013.02.001} {\bibfield  {journal}
  {\bibinfo  {journal} {Physics Reports}\ }\textbf {\bibinfo {volume} {528}},\
  \bibinfo {pages} {1 } (\bibinfo {year} {2013})}\BibitemShut {NoStop}%
\bibitem [{\citenamefont {Weber}\ \emph {et~al.}(2010)\citenamefont {Weber},
  \citenamefont {Koehl}, \citenamefont {Varley}, \citenamefont {Janotti},
  \citenamefont {Buckley}, \citenamefont {Van~de Walle},\ and\ \citenamefont
  {Awschalom}}]{Weber11052010}%
  \BibitemOpen
  \bibfield  {author} {\bibinfo {author} {\bibfnamefont {J.~R.}\ \bibnamefont
  {Weber}}, \bibinfo {author} {\bibfnamefont {W.~F.}\ \bibnamefont {Koehl}},
  \bibinfo {author} {\bibfnamefont {J.~B.}\ \bibnamefont {Varley}}, \bibinfo
  {author} {\bibfnamefont {A.}~\bibnamefont {Janotti}}, \bibinfo {author}
  {\bibfnamefont {B.~B.}\ \bibnamefont {Buckley}}, \bibinfo {author}
  {\bibfnamefont {C.~G.}\ \bibnamefont {Van~de Walle}}, \ and\ \bibinfo
  {author} {\bibfnamefont {D.~D.}\ \bibnamefont {Awschalom}},\ }\href {\doibase
  10.1073/pnas.1003052107} {\bibfield  {journal} {\bibinfo  {journal}
  {Proceedings of the National Academy of Sciences}\ }\textbf {\bibinfo
  {volume} {107}},\ \bibinfo {pages} {8513} (\bibinfo {year}
  {2010})}\BibitemShut {NoStop}%
\bibitem [{\citenamefont {Koehl}\ \emph {et~al.}(2011)\citenamefont {Koehl},
  \citenamefont {Buckley}, \citenamefont {Heremans}, \citenamefont {Calusine},\
  and\ \citenamefont {Awschalom}}]{KoehlBuckleyHeremansEtAl2011}%
  \BibitemOpen
  \bibfield  {author} {\bibinfo {author} {\bibfnamefont {W.~F.}\ \bibnamefont
  {Koehl}}, \bibinfo {author} {\bibfnamefont {B.~B.}\ \bibnamefont {Buckley}},
  \bibinfo {author} {\bibfnamefont {F.~J.}\ \bibnamefont {Heremans}}, \bibinfo
  {author} {\bibfnamefont {G.}~\bibnamefont {Calusine}}, \ and\ \bibinfo
  {author} {\bibfnamefont {D.~D.}\ \bibnamefont {Awschalom}},\ }\href
  {http://dx.doi.org/10.1038/nature10562} {\bibfield  {journal} {\bibinfo
  {journal} {Nature}\ }\textbf {\bibinfo {volume} {479}},\ \bibinfo {pages}
  {84} (\bibinfo {year} {2011})}\BibitemShut {NoStop}%
\bibitem [{\citenamefont {Geim}\ and\ \citenamefont
  {Grigorieva}(2013)}]{GeimGrigorieva2013}%
  \BibitemOpen
  \bibfield  {author} {\bibinfo {author} {\bibfnamefont {A.~K.}\ \bibnamefont
  {Geim}}\ and\ \bibinfo {author} {\bibfnamefont {I.~V.}\ \bibnamefont
  {Grigorieva}},\ }\href {http://dx.doi.org/10.1038/nature12385} {\bibfield
  {journal} {\bibinfo  {journal} {Nature}\ }\textbf {\bibinfo {volume} {499}},\
  \bibinfo {pages} {419} (\bibinfo {year} {2013})}\BibitemShut {NoStop}%
\bibitem [{\citenamefont {Wrachtrup}(2016)}]{Wrachtrup2016}%
  \BibitemOpen
  \bibfield  {author} {\bibinfo {author} {\bibfnamefont {J.}~\bibnamefont
  {Wrachtrup}},\ }\href {http://dx.doi.org/10.1038/nnano.2015.265} {\bibfield
  {journal} {\bibinfo  {journal} {Nat. Nanotechnol.}\ }\textbf {\bibinfo
  {volume} {11}},\ \bibinfo {pages} {7} (\bibinfo {year} {2016})}\BibitemShut
  {NoStop}%
\bibitem [{\citenamefont {Cassabois}\ \emph {et~al.}(2016)\citenamefont
  {Cassabois}, \citenamefont {Valvin},\ and\ \citenamefont
  {Gil}}]{Cassabois2016}%
  \BibitemOpen
  \bibfield  {author} {\bibinfo {author} {\bibfnamefont {G.}~\bibnamefont
  {Cassabois}}, \bibinfo {author} {\bibfnamefont {P.}~\bibnamefont {Valvin}}, \
  and\ \bibinfo {author} {\bibfnamefont {B.}~\bibnamefont {Gil}},\ }\href
  {http://dx.doi.org/10.1038/nphoton.2015.277} {\bibfield  {journal} {\bibinfo
  {journal} {Nature Photonics}\ }\textbf {\bibinfo {volume} {10}},\ \bibinfo
  {pages} {262} (\bibinfo {year} {2016})}\BibitemShut {NoStop}%
\bibitem [{\citenamefont {Dai}\ \emph {et~al.}(2015)\citenamefont {Dai},
  \citenamefont {Ma}, \citenamefont {Liu}, \citenamefont {Andersen},
  \citenamefont {Fei}, \citenamefont {Goldflam}, \citenamefont {Wagner},
  \citenamefont {Watanabe}, \citenamefont {Taniguchi}, \citenamefont
  {Thiemens}, \citenamefont {Keilmann}, \citenamefont {Janssen}, \citenamefont
  {Zhu}, \citenamefont {Jarillo-Herrero}, \citenamefont {Fogler},\ and\
  \citenamefont {Basov}}]{Dai2015}%
  \BibitemOpen
  \bibfield  {author} {\bibinfo {author} {\bibfnamefont {S.}~\bibnamefont
  {Dai}}, \bibinfo {author} {\bibfnamefont {Q.}~\bibnamefont {Ma}}, \bibinfo
  {author} {\bibfnamefont {M.~K.}\ \bibnamefont {Liu}}, \bibinfo {author}
  {\bibfnamefont {T.}~\bibnamefont {Andersen}}, \bibinfo {author}
  {\bibfnamefont {Z.}~\bibnamefont {Fei}}, \bibinfo {author} {\bibfnamefont
  {M.~D.}\ \bibnamefont {Goldflam}}, \bibinfo {author} {\bibfnamefont
  {M.}~\bibnamefont {Wagner}}, \bibinfo {author} {\bibfnamefont
  {K.}~\bibnamefont {Watanabe}}, \bibinfo {author} {\bibfnamefont
  {T.}~\bibnamefont {Taniguchi}}, \bibinfo {author} {\bibfnamefont
  {M.}~\bibnamefont {Thiemens}}, \bibinfo {author} {\bibfnamefont
  {F.}~\bibnamefont {Keilmann}}, \bibinfo {author} {\bibfnamefont {G.~C.
  A.~M.}\ \bibnamefont {Janssen}}, \bibinfo {author} {\bibfnamefont {S.-E.}\
  \bibnamefont {Zhu}}, \bibinfo {author} {\bibfnamefont {P.}~\bibnamefont
  {Jarillo-Herrero}}, \bibinfo {author} {\bibfnamefont {M.~M.}\ \bibnamefont
  {Fogler}}, \ and\ \bibinfo {author} {\bibfnamefont {D.~N.}\ \bibnamefont
  {Basov}},\ }\href {http://dx.doi.org/10.1038/nnano.2015.131} {\bibfield
  {journal} {\bibinfo  {journal} {Nat. Nanotechnol.}\ }\textbf {\bibinfo
  {volume} {10}},\ \bibinfo {pages} {682} (\bibinfo {year} {2015})}\BibitemShut
  {NoStop}%
\bibitem [{\citenamefont {Michel}\ and\ \citenamefont
  {Verberck}(2009)}]{PhysRevB.80.224301}%
  \BibitemOpen
  \bibfield  {author} {\bibinfo {author} {\bibfnamefont {K.~H.}\ \bibnamefont
  {Michel}}\ and\ \bibinfo {author} {\bibfnamefont {B.}~\bibnamefont
  {Verberck}},\ }\href {\doibase 10.1103/PhysRevB.80.224301} {\bibfield
  {journal} {\bibinfo  {journal} {Phys. Rev. B}\ }\textbf {\bibinfo {volume}
  {80}},\ \bibinfo {pages} {224301} (\bibinfo {year} {2009})}\BibitemShut
  {NoStop}%
\bibitem [{\citenamefont {Michel}\ and\ \citenamefont
  {Verberck}(2011)}]{PhysRevB.83.115328}%
  \BibitemOpen
  \bibfield  {author} {\bibinfo {author} {\bibfnamefont {K.~H.}\ \bibnamefont
  {Michel}}\ and\ \bibinfo {author} {\bibfnamefont {B.}~\bibnamefont
  {Verberck}},\ }\href {\doibase 10.1103/PhysRevB.83.115328} {\bibfield
  {journal} {\bibinfo  {journal} {Phys. Rev. B}\ }\textbf {\bibinfo {volume}
  {83}},\ \bibinfo {pages} {115328} (\bibinfo {year} {2011})}\BibitemShut
  {NoStop}%
\bibitem [{\citenamefont {Dai}\ \emph {et~al.}(2014)\citenamefont {Dai},
  \citenamefont {Fei}, \citenamefont {Ma}, \citenamefont {Rodin}, \citenamefont
  {Wagner}, \citenamefont {McLeod}, \citenamefont {Liu}, \citenamefont
  {Gannett}, \citenamefont {Regan}, \citenamefont {Watanabe}, \citenamefont
  {Taniguchi}, \citenamefont {Thiemens}, \citenamefont {Dominguez},
  \citenamefont {Neto}, \citenamefont {Zettl}, \citenamefont {Keilmann},
  \citenamefont {Jarillo-Herrero}, \citenamefont {Fogler},\ and\ \citenamefont
  {Basov}}]{Dai2014}%
  \BibitemOpen
  \bibfield  {author} {\bibinfo {author} {\bibfnamefont {S.}~\bibnamefont
  {Dai}}, \bibinfo {author} {\bibfnamefont {Z.}~\bibnamefont {Fei}}, \bibinfo
  {author} {\bibfnamefont {Q.}~\bibnamefont {Ma}}, \bibinfo {author}
  {\bibfnamefont {A.~S.}\ \bibnamefont {Rodin}}, \bibinfo {author}
  {\bibfnamefont {M.}~\bibnamefont {Wagner}}, \bibinfo {author} {\bibfnamefont
  {A.~S.}\ \bibnamefont {McLeod}}, \bibinfo {author} {\bibfnamefont {M.~K.}\
  \bibnamefont {Liu}}, \bibinfo {author} {\bibfnamefont {W.}~\bibnamefont
  {Gannett}}, \bibinfo {author} {\bibfnamefont {W.}~\bibnamefont {Regan}},
  \bibinfo {author} {\bibfnamefont {K.}~\bibnamefont {Watanabe}}, \bibinfo
  {author} {\bibfnamefont {T.}~\bibnamefont {Taniguchi}}, \bibinfo {author}
  {\bibfnamefont {M.}~\bibnamefont {Thiemens}}, \bibinfo {author}
  {\bibfnamefont {G.}~\bibnamefont {Dominguez}}, \bibinfo {author}
  {\bibfnamefont {A.~H.~C.}\ \bibnamefont {Neto}}, \bibinfo {author}
  {\bibfnamefont {A.}~\bibnamefont {Zettl}}, \bibinfo {author} {\bibfnamefont
  {F.}~\bibnamefont {Keilmann}}, \bibinfo {author} {\bibfnamefont
  {P.}~\bibnamefont {Jarillo-Herrero}}, \bibinfo {author} {\bibfnamefont
  {M.~M.}\ \bibnamefont {Fogler}}, \ and\ \bibinfo {author} {\bibfnamefont
  {D.~N.}\ \bibnamefont {Basov}},\ }\href {\doibase 10.1126/science.1246833}
  {\bibfield  {journal} {\bibinfo  {journal} {Science}\ }\textbf {\bibinfo
  {volume} {343}},\ \bibinfo {pages} {1125} (\bibinfo {year}
  {2014})}\BibitemShut {NoStop}%
\bibitem [{\citenamefont {Wong}\ \emph {et~al.}(2015)\citenamefont {Wong},
  \citenamefont {Velasco~Jr}, \citenamefont {Ju}, \citenamefont {Lee},
  \citenamefont {Kahn}, \citenamefont {Tsai}, \citenamefont {Germany},
  \citenamefont {Taniguchi}, \citenamefont {Watanabe}, \citenamefont {Zettl},
  \citenamefont {Wang},\ and\ \citenamefont {Crommie}}]{Wong2015}%
  \BibitemOpen
  \bibfield  {author} {\bibinfo {author} {\bibfnamefont {D.}~\bibnamefont
  {Wong}}, \bibinfo {author} {\bibfnamefont {J.}~\bibnamefont {Velasco~Jr}},
  \bibinfo {author} {\bibfnamefont {L.}~\bibnamefont {Ju}}, \bibinfo {author}
  {\bibfnamefont {J.}~\bibnamefont {Lee}}, \bibinfo {author} {\bibfnamefont
  {S.}~\bibnamefont {Kahn}}, \bibinfo {author} {\bibfnamefont {H.-Z.}\
  \bibnamefont {Tsai}}, \bibinfo {author} {\bibfnamefont {C.}~\bibnamefont
  {Germany}}, \bibinfo {author} {\bibfnamefont {T.}~\bibnamefont {Taniguchi}},
  \bibinfo {author} {\bibfnamefont {K.}~\bibnamefont {Watanabe}}, \bibinfo
  {author} {\bibfnamefont {A.}~\bibnamefont {Zettl}}, \bibinfo {author}
  {\bibfnamefont {F.}~\bibnamefont {Wang}}, \ and\ \bibinfo {author}
  {\bibfnamefont {M.~F.}\ \bibnamefont {Crommie}},\ }\href
  {http://dx.doi.org/10.1038/nnano.2015.188} {\bibfield  {journal} {\bibinfo
  {journal} {Nat. Nanotechnol.}\ }\textbf {\bibinfo {volume} {10}},\ \bibinfo
  {pages} {949} (\bibinfo {year} {2015})}\BibitemShut {NoStop}%
\bibitem [{\citenamefont {Museur}\ \emph {et~al.}(2008)\citenamefont {Museur},
  \citenamefont {Feldbach},\ and\ \citenamefont {Kanaev}}]{PhysRevB.78.155204}%
  \BibitemOpen
  \bibfield  {author} {\bibinfo {author} {\bibfnamefont {L.}~\bibnamefont
  {Museur}}, \bibinfo {author} {\bibfnamefont {E.}~\bibnamefont {Feldbach}}, \
  and\ \bibinfo {author} {\bibfnamefont {A.}~\bibnamefont {Kanaev}},\ }\href
  {\doibase 10.1103/PhysRevB.78.155204} {\bibfield  {journal} {\bibinfo
  {journal} {Phys. Rev. B}\ }\textbf {\bibinfo {volume} {78}},\ \bibinfo
  {pages} {155204} (\bibinfo {year} {2008})}\BibitemShut {NoStop}%
\bibitem [{\citenamefont {Meuret}\ \emph {et~al.}(2015)\citenamefont {Meuret},
  \citenamefont {Tizei}, \citenamefont {Cazimajou}, \citenamefont
  {Bourrellier}, \citenamefont {Chang}, \citenamefont {Treussart},\ and\
  \citenamefont {Kociak}}]{Meuret2015}%
  \BibitemOpen
  \bibfield  {author} {\bibinfo {author} {\bibfnamefont {S.}~\bibnamefont
  {Meuret}}, \bibinfo {author} {\bibfnamefont {L.~H.~G.}\ \bibnamefont
  {Tizei}}, \bibinfo {author} {\bibfnamefont {T.}~\bibnamefont {Cazimajou}},
  \bibinfo {author} {\bibfnamefont {R.}~\bibnamefont {Bourrellier}}, \bibinfo
  {author} {\bibfnamefont {H.~C.}\ \bibnamefont {Chang}}, \bibinfo {author}
  {\bibfnamefont {F.}~\bibnamefont {Treussart}}, \ and\ \bibinfo {author}
  {\bibfnamefont {M.}~\bibnamefont {Kociak}},\ }\href
  {http://journals.aps.org/prl/accepted/26073Y27Mad14748f4f97735f284a60187b87491f}
  {\bibfield  {journal} {\bibinfo  {journal} {Phys. Rev. Lett.}\ }\textbf
  {\bibinfo {volume} {114}},\ \bibinfo {pages} {197401} (\bibinfo {year}
  {2015})}\BibitemShut {NoStop}%
\bibitem [{\citenamefont {Bourrellier}\ \emph {et~al.}(2016)\citenamefont
  {Bourrellier}, \citenamefont {Meuret}, \citenamefont {Tararan}, \citenamefont
  {StÃ©phan}, \citenamefont {Kociak}, \citenamefont {Tizei},\ and\
  \citenamefont {Zobelli}}]{Bourrellier2016}%
  \BibitemOpen
  \bibfield  {author} {\bibinfo {author} {\bibfnamefont {R.}~\bibnamefont
  {Bourrellier}}, \bibinfo {author} {\bibfnamefont {S.}~\bibnamefont {Meuret}},
  \bibinfo {author} {\bibfnamefont {A.}~\bibnamefont {Tararan}}, \bibinfo
  {author} {\bibfnamefont {O.}~\bibnamefont {StÃ©phan}}, \bibinfo {author}
  {\bibfnamefont {M.}~\bibnamefont {Kociak}}, \bibinfo {author} {\bibfnamefont
  {L.~H.~G.}\ \bibnamefont {Tizei}}, \ and\ \bibinfo {author} {\bibfnamefont
  {A.}~\bibnamefont {Zobelli}},\ }\href {\doibase 10.1021/acs.nanolett.6b01368}
  {\bibfield  {journal} {\bibinfo  {journal} {Nano Lett.}\ }\textbf {\bibinfo
  {volume} {16}},\ \bibinfo {pages} {4317} (\bibinfo {year}
  {2016})}\BibitemShut {NoStop}%
\bibitem [{\citenamefont {Tran}\ \emph
  {et~al.}(2016{\natexlab{a}})\citenamefont {Tran}, \citenamefont {Bray},
  \citenamefont {Ford}, \citenamefont {Toth},\ and\ \citenamefont
  {Aharonovich}}]{TranBrayFordEtAl2016}%
  \BibitemOpen
  \bibfield  {author} {\bibinfo {author} {\bibfnamefont {T.~T.}\ \bibnamefont
  {Tran}}, \bibinfo {author} {\bibfnamefont {K.}~\bibnamefont {Bray}}, \bibinfo
  {author} {\bibfnamefont {M.~J.}\ \bibnamefont {Ford}}, \bibinfo {author}
  {\bibfnamefont {M.}~\bibnamefont {Toth}}, \ and\ \bibinfo {author}
  {\bibfnamefont {I.}~\bibnamefont {Aharonovich}},\ }\href
  {http://dx.doi.org/10.1038/nnano.2015.242} {\bibfield  {journal} {\bibinfo
  {journal} {Nat. Nanotechnol.}\ }\textbf {\bibinfo {volume} {11}},\ \bibinfo
  {pages} {37} (\bibinfo {year} {2016}{\natexlab{a}})}\BibitemShut {NoStop}%
\bibitem [{\citenamefont {Tran}\ \emph
  {et~al.}(2016{\natexlab{b}})\citenamefont {Tran}, \citenamefont {Zachreson},
  \citenamefont {Berhane}, \citenamefont {Bray}, \citenamefont {Sandstrom},
  \citenamefont {Li}, \citenamefont {Taniguchi}, \citenamefont {Watanabe},
  \citenamefont {Aharonovich},\ and\ \citenamefont
  {Toth}}]{PhysRevApplied.5.034005}%
  \BibitemOpen
  \bibfield  {author} {\bibinfo {author} {\bibfnamefont {T.~T.}\ \bibnamefont
  {Tran}}, \bibinfo {author} {\bibfnamefont {C.}~\bibnamefont {Zachreson}},
  \bibinfo {author} {\bibfnamefont {A.~M.}\ \bibnamefont {Berhane}}, \bibinfo
  {author} {\bibfnamefont {K.}~\bibnamefont {Bray}}, \bibinfo {author}
  {\bibfnamefont {R.~G.}\ \bibnamefont {Sandstrom}}, \bibinfo {author}
  {\bibfnamefont {L.~H.}\ \bibnamefont {Li}}, \bibinfo {author} {\bibfnamefont
  {T.}~\bibnamefont {Taniguchi}}, \bibinfo {author} {\bibfnamefont
  {K.}~\bibnamefont {Watanabe}}, \bibinfo {author} {\bibfnamefont
  {I.}~\bibnamefont {Aharonovich}}, \ and\ \bibinfo {author} {\bibfnamefont
  {M.}~\bibnamefont {Toth}},\ }\href {\doibase 10.1103/PhysRevApplied.5.034005}
  {\bibfield  {journal} {\bibinfo  {journal} {Phys. Rev. Applied}\ }\textbf
  {\bibinfo {volume} {5}},\ \bibinfo {pages} {034005} (\bibinfo {year}
  {2016}{\natexlab{b}})}\BibitemShut {NoStop}%
\bibitem [{\citenamefont {Schell}\ \emph {et~al.}(2016)\citenamefont {Schell},
  \citenamefont {Tran}, \citenamefont {Takashima}, \citenamefont {Takeuchi},\
  and\ \citenamefont {Aharonovich}}]{2016arXiv160609364S}%
  \BibitemOpen
  \bibfield  {author} {\bibinfo {author} {\bibfnamefont {A.~W.}\ \bibnamefont
  {Schell}}, \bibinfo {author} {\bibfnamefont {T.~T.}\ \bibnamefont {Tran}},
  \bibinfo {author} {\bibfnamefont {H.}~\bibnamefont {Takashima}}, \bibinfo
  {author} {\bibfnamefont {S.}~\bibnamefont {Takeuchi}}, \ and\ \bibinfo
  {author} {\bibfnamefont {I.}~\bibnamefont {Aharonovich}},\ }\href
  {http://scitation.aip.org/content/aip/journal/app/1/9/10.1063/1.4961684}
  {\bibfield  {journal} {\bibinfo  {journal} {APL Photonics}\ }\textbf
  {\bibinfo {volume} {1}},\ \bibinfo {eid} {091302} (\bibinfo {year}
  {2016})}\BibitemShut {NoStop}%
\bibitem [{\citenamefont {{Mart{\'{\i}}nez}}\ \emph {et~al.}(2016)\citenamefont
  {{Mart{\'{\i}}nez}}, \citenamefont {{Pelini}}, \citenamefont {{Waselowski}},
  \citenamefont {{Maze}}, \citenamefont {{Gil}}, \citenamefont {{Cassabois}},\
  and\ \citenamefont {{Jacques}}}]{2016arXiv160604124M}%
  \BibitemOpen
  \bibfield  {author} {\bibinfo {author} {\bibfnamefont {L.~J.}\ \bibnamefont
  {{Mart{\'{\i}}nez}}}, \bibinfo {author} {\bibfnamefont {T.}~\bibnamefont
  {{Pelini}}}, \bibinfo {author} {\bibfnamefont {V.}~\bibnamefont
  {{Waselowski}}}, \bibinfo {author} {\bibfnamefont {J.~R.}\ \bibnamefont
  {{Maze}}}, \bibinfo {author} {\bibfnamefont {B.}~\bibnamefont {{Gil}}},
  \bibinfo {author} {\bibfnamefont {G.}~\bibnamefont {{Cassabois}}}, \ and\
  \bibinfo {author} {\bibfnamefont {V.}~\bibnamefont {{Jacques}}},\ }\href
  {http://arxiv.org/abs/1606.04124} {\bibfield  {journal} {\bibinfo  {journal}
  {ArXiv e-prints}\ } (\bibinfo {year} {2016})},\ \Eprint
  {http://arxiv.org/abs/1606.04124} {arXiv:1606.04124 [cond-mat.mtrl-sci]}
  \BibitemShut {NoStop}%
\bibitem [{\citenamefont {{Jungwirth}}\ \emph {et~al.}(2016)\citenamefont
  {{Jungwirth}}, \citenamefont {{Calderon}}, \citenamefont {{Ji}},
  \citenamefont {{Spencer}}, \citenamefont {{Flatt{\'e}}},\ and\ \citenamefont
  {{Fuchs}}}]{2016arXiv160504445J}%
  \BibitemOpen
  \bibfield  {author} {\bibinfo {author} {\bibfnamefont {N.~R.}\ \bibnamefont
  {{Jungwirth}}}, \bibinfo {author} {\bibfnamefont {B.}~\bibnamefont
  {{Calderon}}}, \bibinfo {author} {\bibfnamefont {Y.}~\bibnamefont {{Ji}}},
  \bibinfo {author} {\bibfnamefont {M.~G.}\ \bibnamefont {{Spencer}}}, \bibinfo
  {author} {\bibfnamefont {M.~E.}\ \bibnamefont {{Flatt{\'e}}}}, \ and\
  \bibinfo {author} {\bibfnamefont {G.~D.}\ \bibnamefont {{Fuchs}}},\ }\href
  {https://arxiv.org/abs/1605.04445} {\bibfield  {journal} {\bibinfo  {journal}
  {ArXiv e-prints}\ } (\bibinfo {year} {2016})},\ \Eprint
  {http://arxiv.org/abs/1605.04445} {arXiv:1605.04445 [cond-mat.mtrl-sci]}
  \BibitemShut {NoStop}%
\bibitem [{\citenamefont {Pacil\'{e}}\ \emph {et~al.}(2008)\citenamefont
  {Pacil\'{e}}, \citenamefont {Meyer}, \citenamefont {Girit},\ and\
  \citenamefont {Zettl}}]{PacileMeyerGiritEtAl2008}%
  \BibitemOpen
  \bibfield  {author} {\bibinfo {author} {\bibfnamefont {D.}~\bibnamefont
  {Pacil\'{e}}}, \bibinfo {author} {\bibfnamefont {J.~C.}\ \bibnamefont
  {Meyer}}, \bibinfo {author} {\bibfnamefont {{\c{C}}.~O.}\ \bibnamefont
  {Girit}}, \ and\ \bibinfo {author} {\bibfnamefont {A.}~\bibnamefont
  {Zettl}},\ }\href {\doibase http://dx.doi.org/10.1063/1.2903702} {\bibfield
  {journal} {\bibinfo  {journal} {Applied Physics Letters}\ }\textbf {\bibinfo
  {volume} {92}},\ \bibinfo {eid} {133107} (\bibinfo {year}
  {2008})}\BibitemShut {NoStop}%
\bibitem [{\citenamefont {Gorbachev}\ \emph {et~al.}(2011)\citenamefont
  {Gorbachev}, \citenamefont {Riaz}, \citenamefont {Nair}, \citenamefont
  {Jalil}, \citenamefont {Britnell}, \citenamefont {Belle}, \citenamefont
  {Hill}, \citenamefont {Novoselov}, \citenamefont {Watanabe}, \citenamefont
  {Taniguchi}, \citenamefont {Geim},\ and\ \citenamefont
  {Blake}}]{SMLL:SMLL201001628}%
  \BibitemOpen
  \bibfield  {author} {\bibinfo {author} {\bibfnamefont {R.~V.}\ \bibnamefont
  {Gorbachev}}, \bibinfo {author} {\bibfnamefont {I.}~\bibnamefont {Riaz}},
  \bibinfo {author} {\bibfnamefont {R.~R.}\ \bibnamefont {Nair}}, \bibinfo
  {author} {\bibfnamefont {R.}~\bibnamefont {Jalil}}, \bibinfo {author}
  {\bibfnamefont {L.}~\bibnamefont {Britnell}}, \bibinfo {author}
  {\bibfnamefont {B.~D.}\ \bibnamefont {Belle}}, \bibinfo {author}
  {\bibfnamefont {E.~W.}\ \bibnamefont {Hill}}, \bibinfo {author}
  {\bibfnamefont {K.~S.}\ \bibnamefont {Novoselov}}, \bibinfo {author}
  {\bibfnamefont {K.}~\bibnamefont {Watanabe}}, \bibinfo {author}
  {\bibfnamefont {T.}~\bibnamefont {Taniguchi}}, \bibinfo {author}
  {\bibfnamefont {A.~K.}\ \bibnamefont {Geim}}, \ and\ \bibinfo {author}
  {\bibfnamefont {P.}~\bibnamefont {Blake}},\ }\href {\doibase
  10.1002/smll.201001628} {\bibfield  {journal} {\bibinfo  {journal} {Small}\
  }\textbf {\bibinfo {volume} {7}},\ \bibinfo {pages} {465} (\bibinfo {year}
  {2011})}\BibitemShut {NoStop}%
\bibitem [{\citenamefont {Novoselov}\ \emph {et~al.}(2005)\citenamefont
  {Novoselov}, \citenamefont {Jiang}, \citenamefont {Schedin}, \citenamefont
  {Booth}, \citenamefont {Khotkevich}, \citenamefont {Morozov},\ and\
  \citenamefont {Geim}}]{Novoselov26072005}%
  \BibitemOpen
  \bibfield  {author} {\bibinfo {author} {\bibfnamefont {K.~S.}\ \bibnamefont
  {Novoselov}}, \bibinfo {author} {\bibfnamefont {D.}~\bibnamefont {Jiang}},
  \bibinfo {author} {\bibfnamefont {F.}~\bibnamefont {Schedin}}, \bibinfo
  {author} {\bibfnamefont {T.~J.}\ \bibnamefont {Booth}}, \bibinfo {author}
  {\bibfnamefont {V.~V.}\ \bibnamefont {Khotkevich}}, \bibinfo {author}
  {\bibfnamefont {S.~V.}\ \bibnamefont {Morozov}}, \ and\ \bibinfo {author}
  {\bibfnamefont {A.~K.}\ \bibnamefont {Geim}},\ }\href {\doibase
  10.1073/pnas.0502848102} {\bibfield  {journal} {\bibinfo  {journal} {Proc.
  Natl. Acad. Sci. U.S.A.}\ }\textbf {\bibinfo {volume} {102}},\ \bibinfo
  {pages} {10451} (\bibinfo {year} {2005})}\BibitemShut {NoStop}%
\bibitem [{\citenamefont {Reich}\ \emph {et~al.}(2005)\citenamefont {Reich},
  \citenamefont {Ferrari}, \citenamefont {Arenal}, \citenamefont {Loiseau},
  \citenamefont {Bello},\ and\ \citenamefont {Robertson}}]{PhysRevB.71.205201}%
  \BibitemOpen
  \bibfield  {author} {\bibinfo {author} {\bibfnamefont {S.}~\bibnamefont
  {Reich}}, \bibinfo {author} {\bibfnamefont {A.~C.}\ \bibnamefont {Ferrari}},
  \bibinfo {author} {\bibfnamefont {R.}~\bibnamefont {Arenal}}, \bibinfo
  {author} {\bibfnamefont {A.}~\bibnamefont {Loiseau}}, \bibinfo {author}
  {\bibfnamefont {I.}~\bibnamefont {Bello}}, \ and\ \bibinfo {author}
  {\bibfnamefont {J.}~\bibnamefont {Robertson}},\ }\href {\doibase
  10.1103/PhysRevB.71.205201} {\bibfield  {journal} {\bibinfo  {journal} {Phys.
  Rev. B}\ }\textbf {\bibinfo {volume} {71}},\ \bibinfo {pages} {205201}
  (\bibinfo {year} {2005})}\BibitemShut {NoStop}%
\bibitem [{Sup()}]{Supplement}%
  \BibitemOpen
  \href@noop {} {\ }\bibinfo {note} {See the supporting information online for
  further details.}\BibitemShut {Stop}%
\bibitem [{\citenamefont {Maradudin}(1966)}]{Maradudin1966}%
  \BibitemOpen
  \bibfield  {author} {\bibinfo {author} {\bibfnamefont {A.}~\bibnamefont
  {Maradudin}}\ }(\bibinfo  {publisher} {Academic Press},\ \bibinfo {year}
  {1966})\ pp.\ \bibinfo {pages} {273 -- 420}\BibitemShut {NoStop}%
\bibitem [{\citenamefont {Davies}(1974)}]{Davies1974}%
  \BibitemOpen
  \bibfield  {author} {\bibinfo {author} {\bibfnamefont {G.}~\bibnamefont
  {Davies}},\ }\href {http://stacks.iop.org/0022-3719/7/i=20/a=019} {\bibfield
  {journal} {\bibinfo  {journal} {Journal of Physics C: Solid State Physics}\
  }\textbf {\bibinfo {volume} {7}},\ \bibinfo {pages} {3797} (\bibinfo {year}
  {1974})}\BibitemShut {NoStop}%
\bibitem [{\citenamefont {Fu}\ \emph {et~al.}(2009)\citenamefont {Fu},
  \citenamefont {Santori}, \citenamefont {Barclay}, \citenamefont {Rogers},
  \citenamefont {Manson},\ and\ \citenamefont
  {Beausoleil}}]{PhysRevLett.103.256404}%
  \BibitemOpen
  \bibfield  {author} {\bibinfo {author} {\bibfnamefont {K.-M.~C.}\
  \bibnamefont {Fu}}, \bibinfo {author} {\bibfnamefont {C.}~\bibnamefont
  {Santori}}, \bibinfo {author} {\bibfnamefont {P.~E.}\ \bibnamefont
  {Barclay}}, \bibinfo {author} {\bibfnamefont {L.~J.}\ \bibnamefont {Rogers}},
  \bibinfo {author} {\bibfnamefont {N.~B.}\ \bibnamefont {Manson}}, \ and\
  \bibinfo {author} {\bibfnamefont {R.~G.}\ \bibnamefont {Beausoleil}},\ }\href
  {\doibase 10.1103/PhysRevLett.103.256404} {\bibfield  {journal} {\bibinfo
  {journal} {Phys. Rev. Lett.}\ }\textbf {\bibinfo {volume} {103}},\ \bibinfo
  {pages} {256404} (\bibinfo {year} {2009})}\BibitemShut {NoStop}%
\bibitem [{\citenamefont {Abtew}\ \emph {et~al.}(2011)\citenamefont {Abtew},
  \citenamefont {Sun}, \citenamefont {Shih}, \citenamefont {Dev}, \citenamefont
  {Zhang},\ and\ \citenamefont {Zhang}}]{PhysRevLett.107.146403}%
  \BibitemOpen
  \bibfield  {author} {\bibinfo {author} {\bibfnamefont {T.~A.}\ \bibnamefont
  {Abtew}}, \bibinfo {author} {\bibfnamefont {Y.~Y.}\ \bibnamefont {Sun}},
  \bibinfo {author} {\bibfnamefont {B.-C.}\ \bibnamefont {Shih}}, \bibinfo
  {author} {\bibfnamefont {P.}~\bibnamefont {Dev}}, \bibinfo {author}
  {\bibfnamefont {S.~B.}\ \bibnamefont {Zhang}}, \ and\ \bibinfo {author}
  {\bibfnamefont {P.}~\bibnamefont {Zhang}},\ }\href {\doibase
  10.1103/PhysRevLett.107.146403} {\bibfield  {journal} {\bibinfo  {journal}
  {Phys. Rev. Lett.}\ }\textbf {\bibinfo {volume} {107}},\ \bibinfo {pages}
  {146403} (\bibinfo {year} {2011})}\BibitemShut {NoStop}%
\bibitem [{\citenamefont {Kitson}\ \emph {et~al.}(1998)\citenamefont {Kitson},
  \citenamefont {Jonsson}, \citenamefont {Rarity},\ and\ \citenamefont
  {Tapster}}]{PhysRevA.58.620}%
  \BibitemOpen
  \bibfield  {author} {\bibinfo {author} {\bibfnamefont {S.~C.}\ \bibnamefont
  {Kitson}}, \bibinfo {author} {\bibfnamefont {P.}~\bibnamefont {Jonsson}},
  \bibinfo {author} {\bibfnamefont {J.~G.}\ \bibnamefont {Rarity}}, \ and\
  \bibinfo {author} {\bibfnamefont {P.~R.}\ \bibnamefont {Tapster}},\ }\href
  {\doibase 10.1103/PhysRevA.58.620} {\bibfield  {journal} {\bibinfo  {journal}
  {Phys. Rev. A}\ }\textbf {\bibinfo {volume} {58}},\ \bibinfo {pages} {620}
  (\bibinfo {year} {1998})}\BibitemShut {NoStop}%
\bibitem [{\citenamefont {Brouri}\ \emph {et~al.}(2000)\citenamefont {Brouri},
  \citenamefont {Beveratos}, \citenamefont {Poizat},\ and\ \citenamefont
  {Grangier}}]{Brouri2000}%
  \BibitemOpen
  \bibfield  {author} {\bibinfo {author} {\bibfnamefont {R.}~\bibnamefont
  {Brouri}}, \bibinfo {author} {\bibfnamefont {A.}~\bibnamefont {Beveratos}},
  \bibinfo {author} {\bibfnamefont {J.-P.}\ \bibnamefont {Poizat}}, \ and\
  \bibinfo {author} {\bibfnamefont {P.}~\bibnamefont {Grangier}},\ }\href
  {\doibase 10.1364/OL.25.001294} {\bibfield  {journal} {\bibinfo  {journal}
  {Opt. Lett.}\ }\textbf {\bibinfo {volume} {25}},\ \bibinfo {pages} {1294}
  (\bibinfo {year} {2000})}\BibitemShut {NoStop}%
\bibitem [{\citenamefont {Lakowicz}(2006)}]{Lakowicz}%
  \BibitemOpen
  \bibfield  {author} {\bibinfo {author} {\bibfnamefont {J.~R.}\ \bibnamefont
  {Lakowicz}},\ }\href {\doibase 10.1007/978-0-387-46312-4} {\emph {\bibinfo
  {title} {Principles of Fluorescence Spectroscopy}}},\ \bibinfo {edition}
  {3rd}\ ed.\ (\bibinfo  {publisher} {Springer US},\ \bibinfo {year}
  {2006})\BibitemShut {NoStop}%
\bibitem [{\citenamefont {Rogers}\ \emph {et~al.}(2014)\citenamefont {Rogers},
  \citenamefont {Jahnke}, \citenamefont {Doherty}, \citenamefont {Dietrich},
  \citenamefont {McGuinness}, \citenamefont {M\"uller}, \citenamefont {Teraji},
  \citenamefont {Sumiya}, \citenamefont {Isoya}, \citenamefont {Manson},\ and\
  \citenamefont {Jelezko}}]{PhysRevB.89.235101}%
  \BibitemOpen
  \bibfield  {author} {\bibinfo {author} {\bibfnamefont {L.~J.}\ \bibnamefont
  {Rogers}}, \bibinfo {author} {\bibfnamefont {K.~D.}\ \bibnamefont {Jahnke}},
  \bibinfo {author} {\bibfnamefont {M.~W.}\ \bibnamefont {Doherty}}, \bibinfo
  {author} {\bibfnamefont {A.}~\bibnamefont {Dietrich}}, \bibinfo {author}
  {\bibfnamefont {L.~P.}\ \bibnamefont {McGuinness}}, \bibinfo {author}
  {\bibfnamefont {C.}~\bibnamefont {M\"uller}}, \bibinfo {author}
  {\bibfnamefont {T.}~\bibnamefont {Teraji}}, \bibinfo {author} {\bibfnamefont
  {H.}~\bibnamefont {Sumiya}}, \bibinfo {author} {\bibfnamefont
  {J.}~\bibnamefont {Isoya}}, \bibinfo {author} {\bibfnamefont {N.~B.}\
  \bibnamefont {Manson}}, \ and\ \bibinfo {author} {\bibfnamefont
  {F.}~\bibnamefont {Jelezko}},\ }\href {\doibase 10.1103/PhysRevB.89.235101}
  {\bibfield  {journal} {\bibinfo  {journal} {Phys. Rev. B}\ }\textbf {\bibinfo
  {volume} {89}},\ \bibinfo {pages} {235101} (\bibinfo {year}
  {2014})}\BibitemShut {NoStop}%
\bibitem [{\citenamefont {Alegre}\ \emph {et~al.}(2007)\citenamefont {Alegre},
  \citenamefont {Santori}, \citenamefont {Medeiros-Ribeiro},\ and\
  \citenamefont {Beausoleil}}]{PhysRevB.76.165205}%
  \BibitemOpen
  \bibfield  {author} {\bibinfo {author} {\bibfnamefont {T.~P.~M.}\
  \bibnamefont {Alegre}}, \bibinfo {author} {\bibfnamefont {C.}~\bibnamefont
  {Santori}}, \bibinfo {author} {\bibfnamefont {G.}~\bibnamefont
  {Medeiros-Ribeiro}}, \ and\ \bibinfo {author} {\bibfnamefont {R.~G.}\
  \bibnamefont {Beausoleil}},\ }\href {\doibase 10.1103/PhysRevB.76.165205}
  {\bibfield  {journal} {\bibinfo  {journal} {Phys. Rev. B}\ }\textbf {\bibinfo
  {volume} {76}},\ \bibinfo {pages} {165205} (\bibinfo {year}
  {2007})}\BibitemShut {NoStop}%
\bibitem [{\citenamefont {Constantinescu}\ \emph {et~al.}(2013)\citenamefont
  {Constantinescu}, \citenamefont {Kuc},\ and\ \citenamefont
  {Heine}}]{PhysRevLett.111.036104}%
  \BibitemOpen
  \bibfield  {author} {\bibinfo {author} {\bibfnamefont {G.}~\bibnamefont
  {Constantinescu}}, \bibinfo {author} {\bibfnamefont {A.}~\bibnamefont {Kuc}},
  \ and\ \bibinfo {author} {\bibfnamefont {T.}~\bibnamefont {Heine}},\ }\href
  {\doibase 10.1103/PhysRevLett.111.036104} {\bibfield  {journal} {\bibinfo
  {journal} {Phys. Rev. Lett.}\ }\textbf {\bibinfo {volume} {111}},\ \bibinfo
  {pages} {036104} (\bibinfo {year} {2013})}\BibitemShut {NoStop}%
\end{thebibliography}


\end{document}

\title{Supplementary Material:  Optical Signatures of Quantum Emitters in Suspended Hexagonal Boron Nitride}% Force line breaks with \\

\author{Annemarie L. Exarhos}
\affiliation{Quantum Engineering Laboratory, Department of Electrical and Systems Engineering, University of Pennsylvania, Philadelphia, Pennsylvania 19104, United States}

\author{David A. Hopper}
\affiliation{Quantum Engineering Laboratory, Department of Electrical and Systems Engineering, University of Pennsylvania, Philadelphia, Pennsylvania 19104, United States}
\affiliation{Department of Physics, University of Pennsylvania, Philadelphia, Pennsylvania 19104, United States}

\author{Richard R. Grote}
\affiliation{Quantum Engineering Laboratory, Department of Electrical and Systems Engineering, University of Pennsylvania, Philadelphia, Pennsylvania 19104, United States}

\author{Audrius Alkauskas}
\affiliation{Center for Physical Sciences and Technology, Vilnius LT-01108, Lithuania}
\affiliation{Department of Physics, Kaunas University of Technology, Kaunas LT-51368, Lithuania}

\author{Lee C. Bassett}
\email{lbassett@seas.upenn.edu}
\affiliation{Quantum Engineering Laboratory, Department of Electrical and Systems Engineering, University of Pennsylvania, Philadelphia, Pennsylvania 19104, United States}

\maketitle

%____________________________________________________
%	SUPPLEMTARY MATERIAL
%____________________________________________________
% define the supplementary material figures
\renewcommand{\thefigure}{S\arabic{figure}}
\renewcommand{\figurename}{Fig.}
\renewcommand{\thesection}{S\arabic{section}}   

\def\experiment{\begin{figure*}[htp]
   \includegraphics{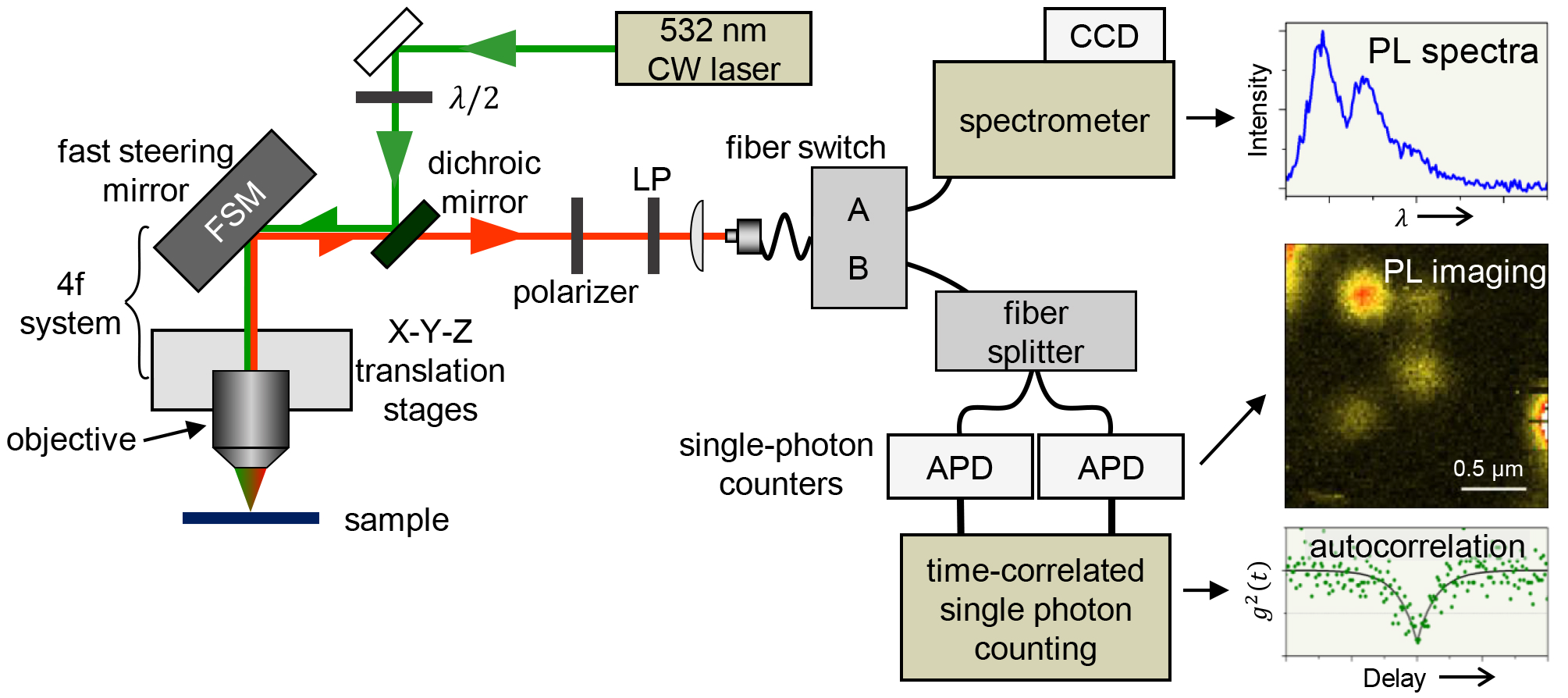}
  \caption{\textbf{Scanning confocal fluorescence microscope.}}
  \label{experiment}
\end{figure*}}

\def\materials{\begin{figure*}[t]
%\fbox{
 \includegraphics[angle=0,scale=1,trim=0.5in 1in 0in 1in,clip=true]{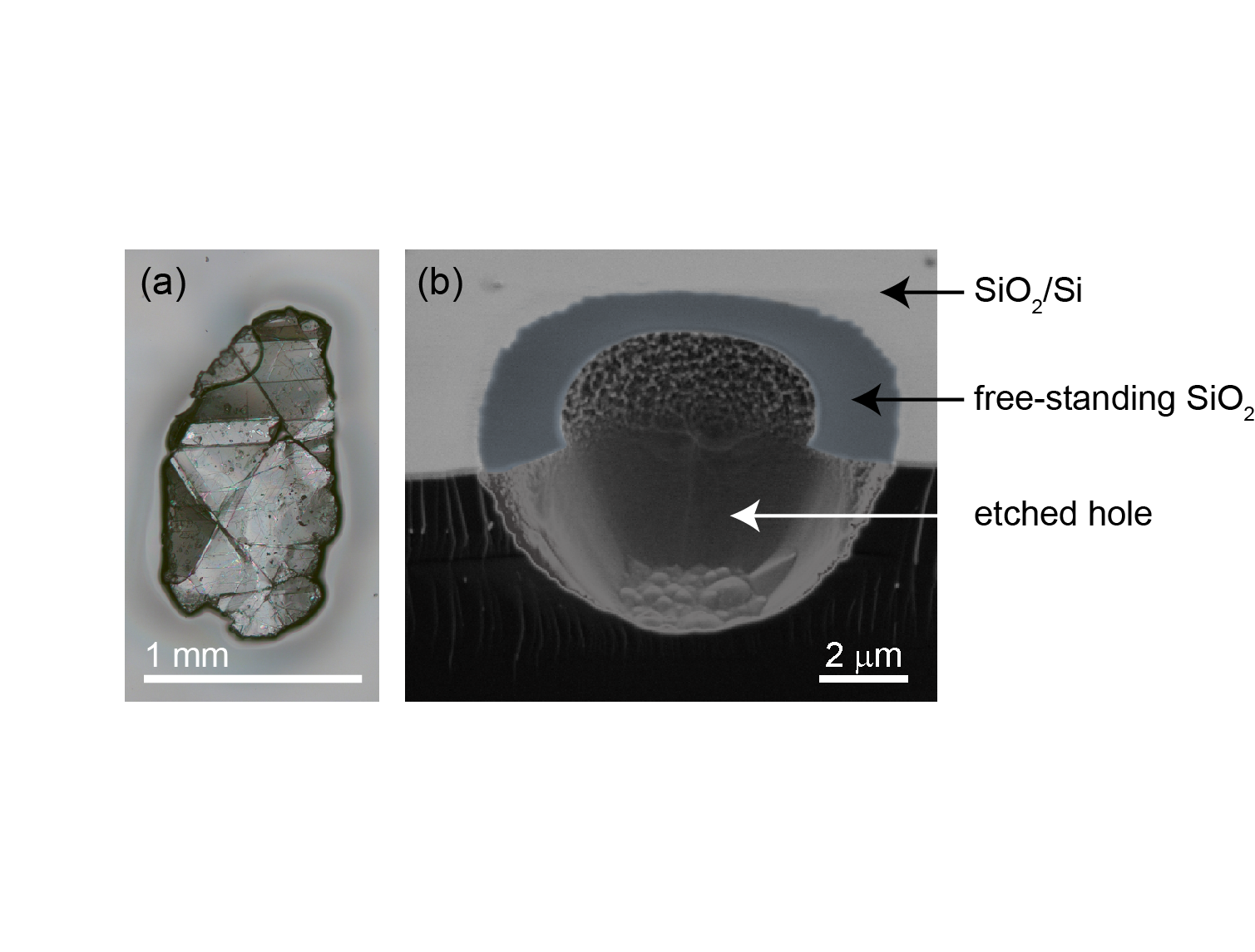}
% }%l, b, r, t
  \caption{\textbf{Sample and substrate used for PL measurements.}  (a) single-crystal h-BN as purchased from HQ Graphene.  (b) SEM image (tilted 52$^{\circ}$ to normal) of the patterned Si/SiO$_2$ substrate indicating the three distinct regions onto which the h-BN is exfoliated.}
  \label{materials}
\end{figure*}}

\def\sample{\begin{figure*}[t]
%\fbox{
 \includegraphics[angle=0,scale=1,trim=0in 0in 0in 0in,clip=true]{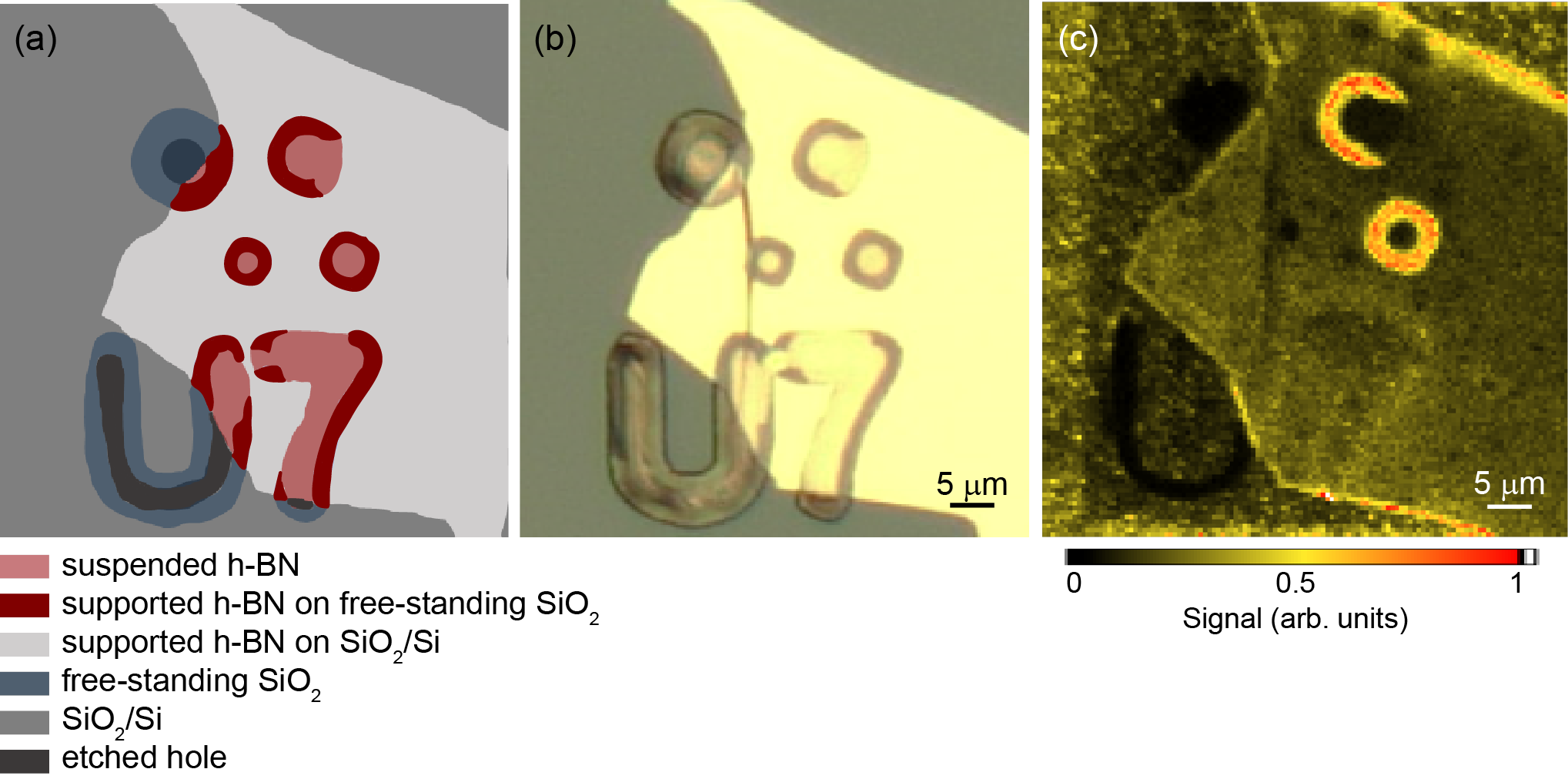}
%}
\caption{\textbf{Representative exfoliated h-BN flake showing different substrate/h-BN regions.}  (a) Cartoon showing 6 distinct regions of the substrate and h-BN.  (b) White light microscope image of an exfoliated h-BN flake on the patterned substrate.  (c) PL image of the same region as in (b).}
  \label{sample}
\end{figure*}}

\def\supportedsuspended{\begin{figure*}[t]
%\fbox{   
\includegraphics[angle=0,scale=1,trim=0in 0in 0in 0in,clip=true]{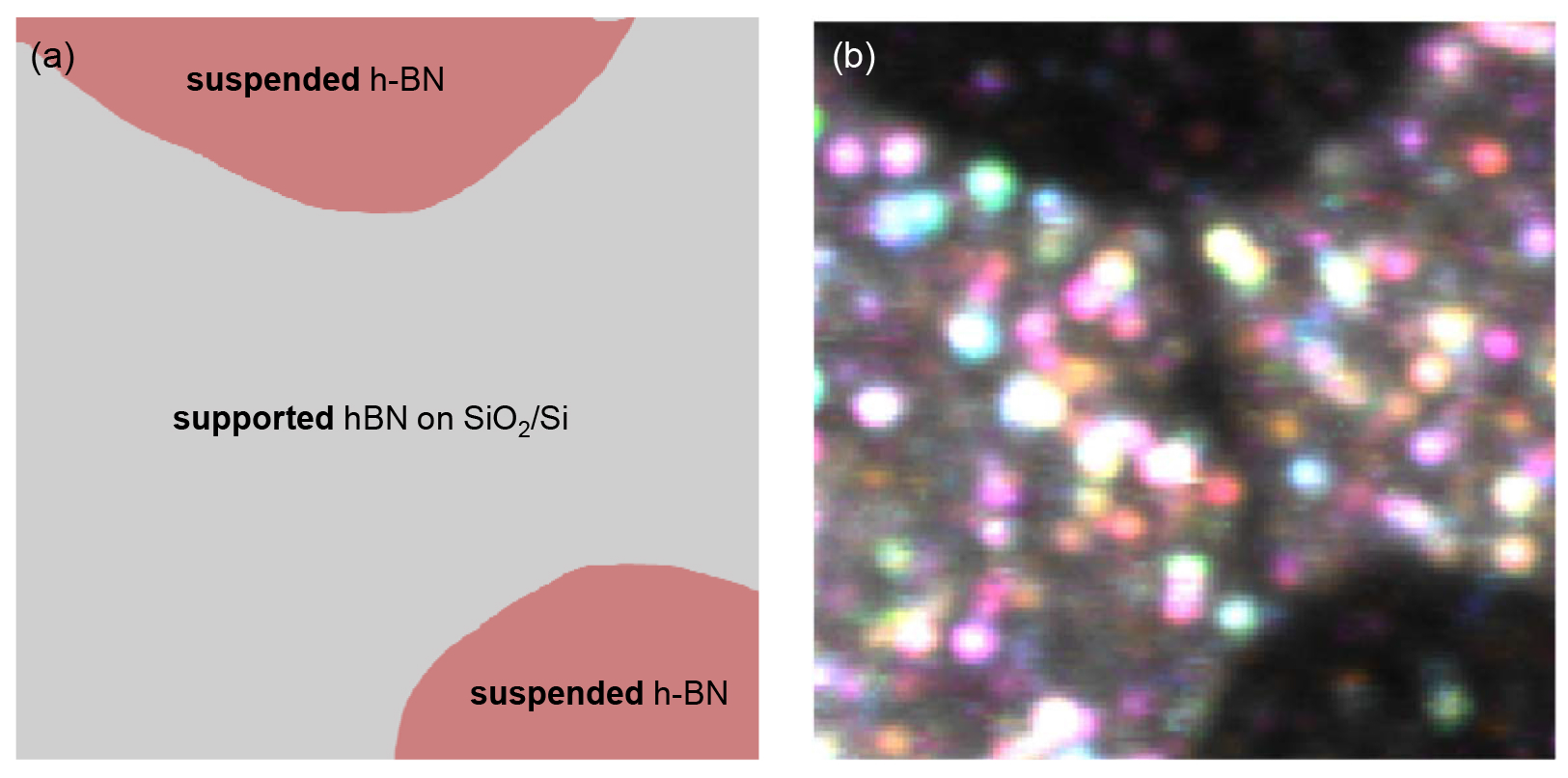}
%}
\caption{\textbf{Comparison of emission from supported and suspended h-BN.}  (a) Cartoon showing supported and suspended h-BN regions.  (b) Polarized PL image of the region defined in (a).  Colors correspond to different orientations of excitation polarization dipoles.}
  \label{supportedsuspended}
\end{figure*}}

\def\Raman{\begin{figure*}[t]
%\fbox{
\includegraphics[angle=0,scale=1,trim=0in 0in 0in 0in,clip=true]{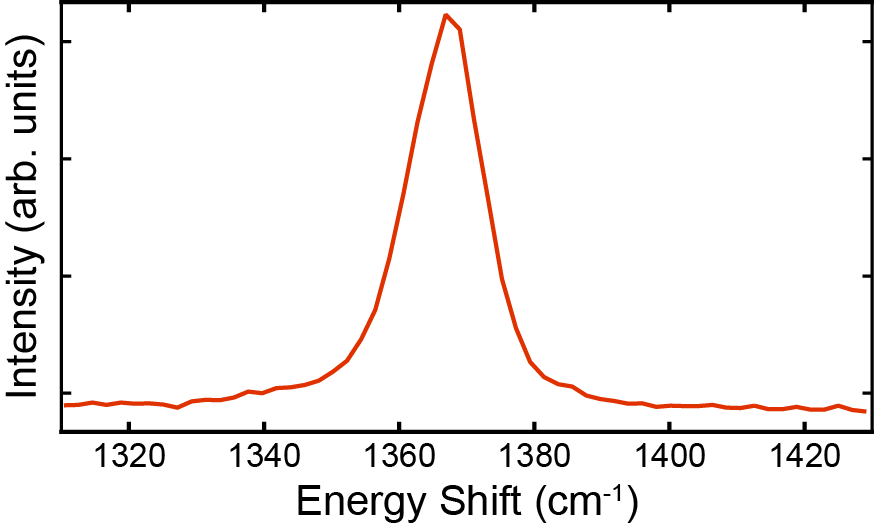}
%}
\caption{\textbf{Raman spectrum of suspended h-BN (from Fig.~1 in the main text).}}
  \label{Raman}
\end{figure*}}

\def\AFM{\begin{figure*}[t]
%\fbox{
 \includegraphics[angle=0,scale=1,trim=0in 0in 0in 0in,clip=true]{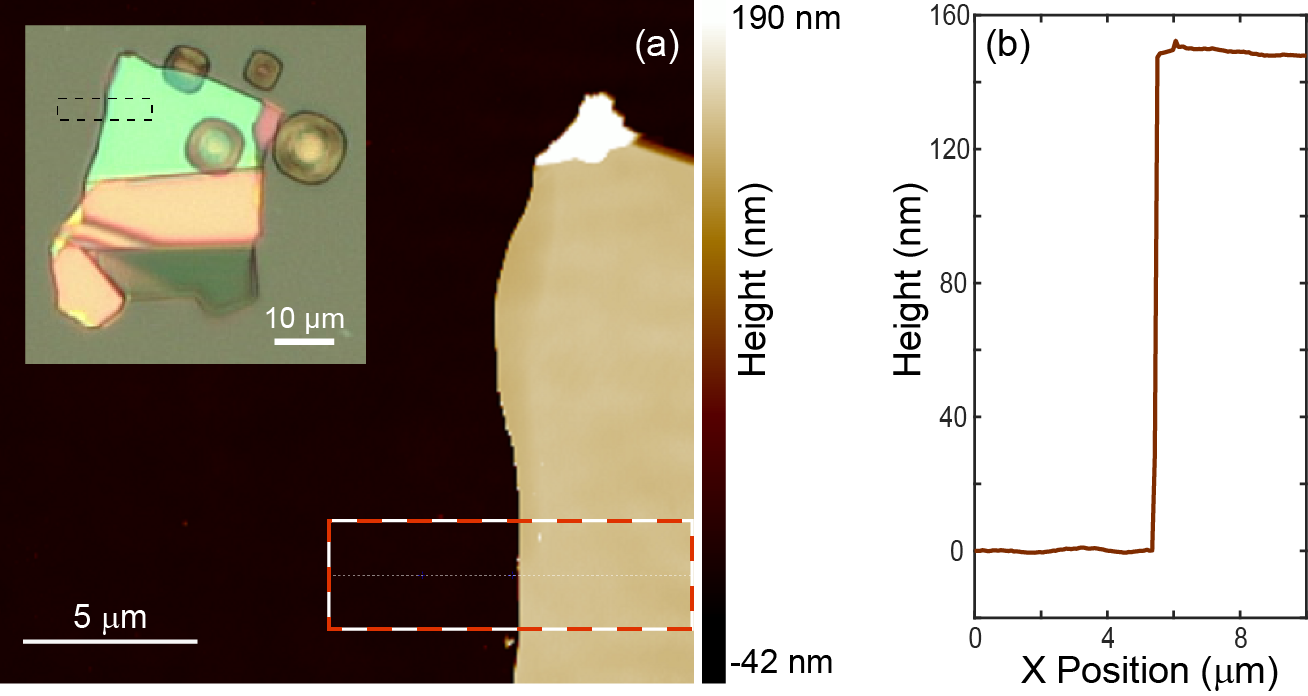}
%}
\caption{\textbf{Atomic force microscopy (AFM) characterization of exfoliated h-BN.}  (a) AFM image of the exfoliated h-BN flake whose defects are studied.  The inset shows an optical image of the entire flake, where defects studied are contained in the partially suspended film near the top of the image.  (b) Height scan for the h-BN studied.  The boxes in the AFM and optical images in (a) denote the region over which the height scan is taken, averaging over the height of the box.}
  \label{AFM}
\end{figure*}}

\def\EBSD{\begin{figure*}[t]
%\fbox{
\includegraphics[angle=0,scale=1,trim=0in 0in 2.29in 1.5in,clip=true]{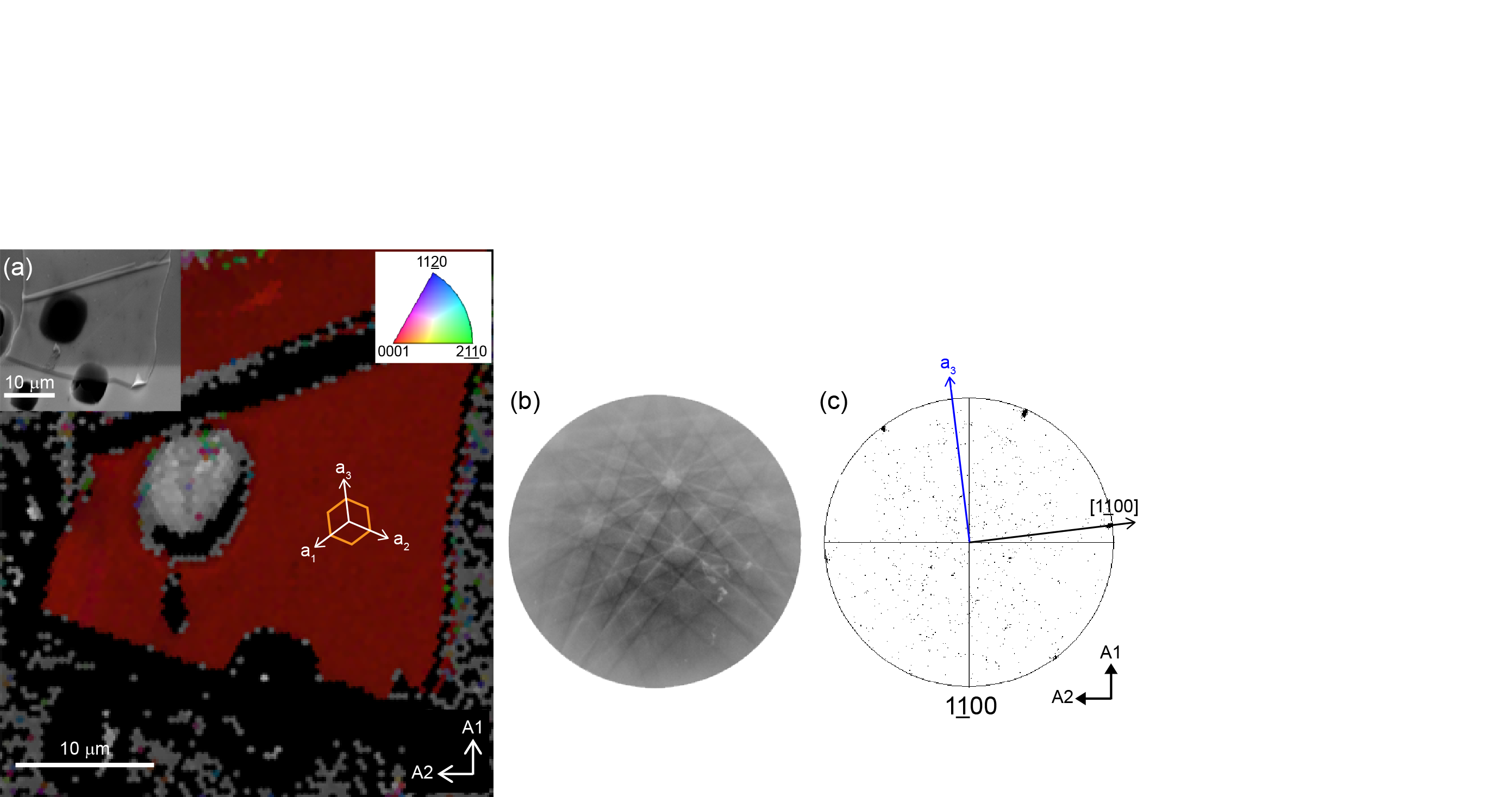}
%} %l, b, r, t
\caption{\textbf{Electron backscatter diffraction (EBSD) on exfoliated h-BN flakes.}  (a) EBSD map of the exfoliated h-BN flake showing the combined image quality (greyscale) and inverse pole figure (color) verifying single-crystallinity.  The inset is an SEM image of the same region.  Sample orientation axes ($\bf{A1}, \bf{A2}$) and lattice vector orientation for the exfoliated h-BN flake ($\bf{a_i}$, $i=1,2,3$) are denoted.  (b) Representative EBSD pattern from the h-BN flake.  (c) $1\bar{1}00$ pole figure showing the in-plane orientation of the h-BN crystal lattice with a representative $\langle 1\bar{1}00\rangle$ and associated lattice vector ($\bf{a_3}$).}
  \label{EBSD}
\end{figure*}}

\def\H3{\begin{figure*}[t]
%\fbox{
 \includegraphics[angle=0,scale=1,trim=0in 0in 0in 0in,clip=true]{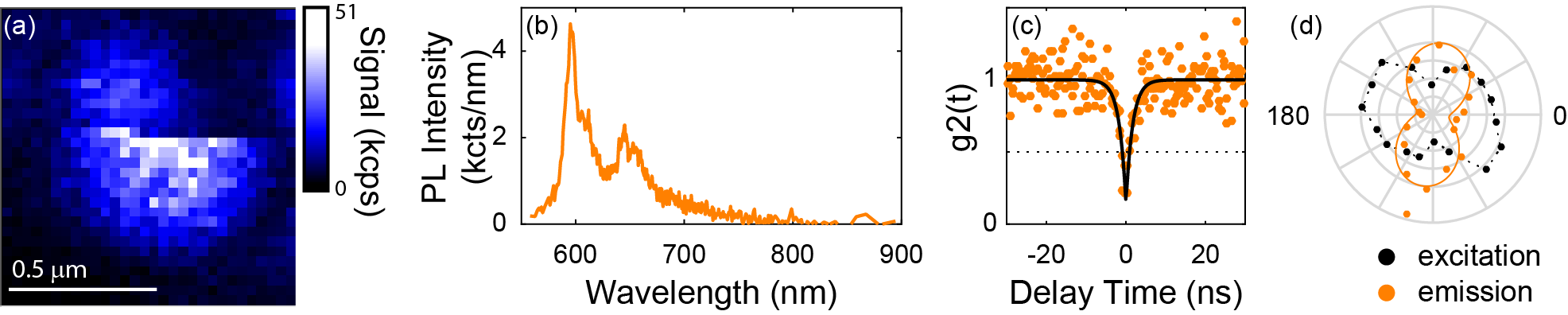}
%}
\caption{\textbf{Single emitter SE5 referenced in the main text.}  (a) PL image, (b) PL spectrum, (c) autocorrelation data and two-level fit, and (d) polarization dependence of the excitation and emission of SE5.}
  \label{H3}
\end{figure*}}

\def\pvalue{\begin{figure*}[t]
%\fbox{
 \includegraphics[angle=0,scale=1,trim=0.1in 0in 0in 0in,clip=true]{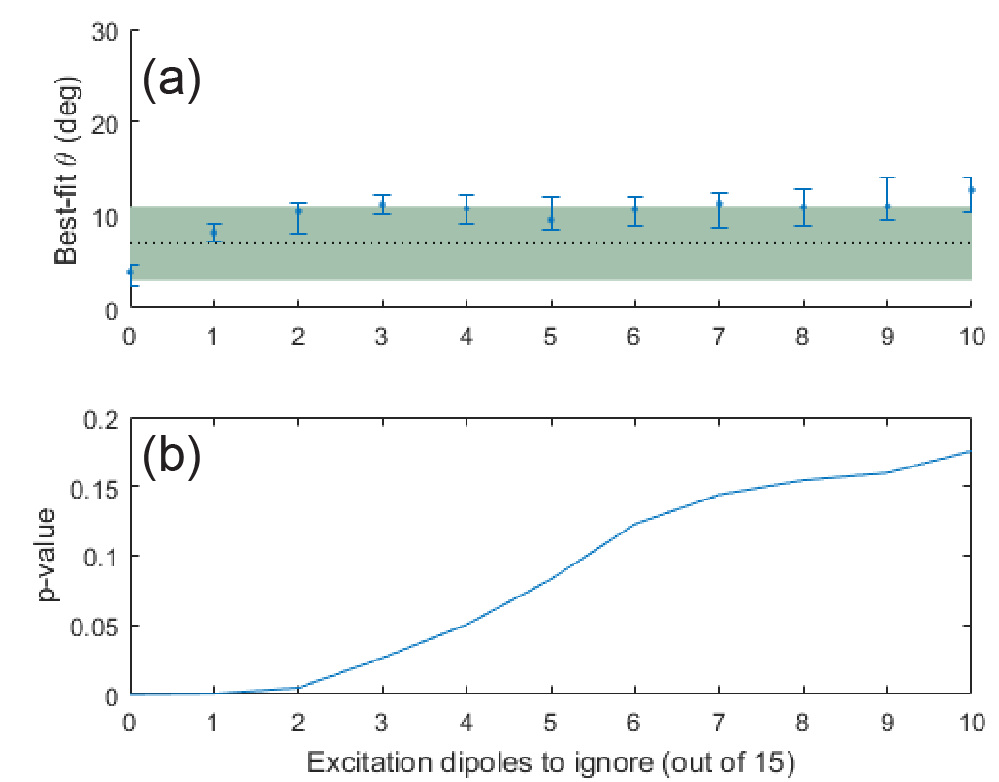}
%}
\caption{\textbf{Statistical analysis of dipole orientation.}  (a) Best-fit crystallographic orientation angle and (b) $p$-value as a function of the number of excitation dipoles ignored in the analysis, $N_{bad}$.  The dotted line and green shaded region indicate the h-BN lattice vector orientation and corresponding uncertainty determined from the EBSD measurement.}
  \label{pvalue}
\end{figure*}}

\def\PSBconstraints{\begin{figure*}[t]
%\fbox{
\includegraphics[angle=0,scale=1,trim=0.05in 0in 0in 0in,clip=true]{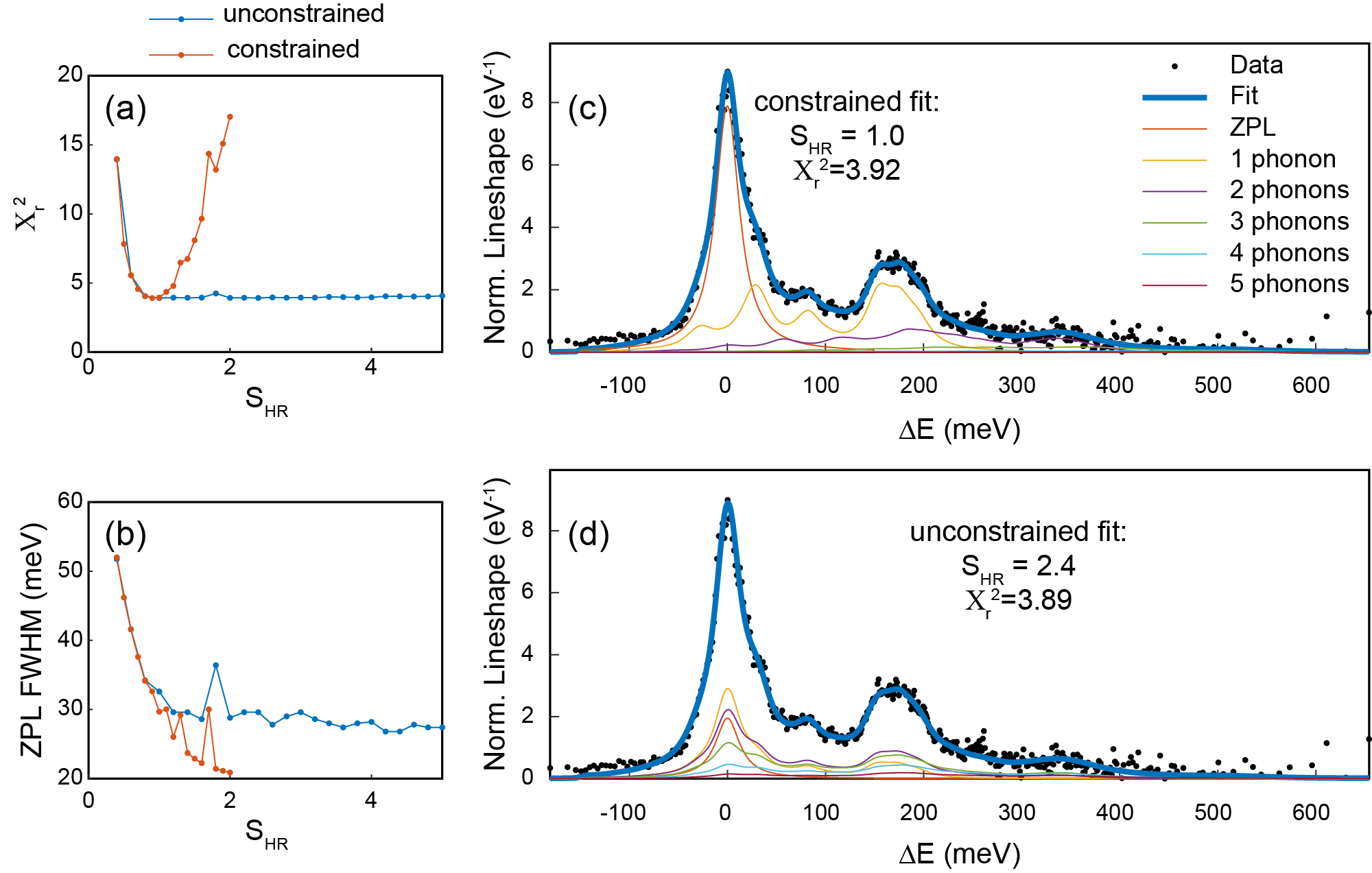}
%}
\caption{\textbf{Constraint effects on low energy phonon contributions to the emission lineshape for SE1.} (a) Reduced chi-squared fit statistic and (b) ZPL width as a function of $S_{HR}$ for constrained and unconstrained fits.  The constraint $S(E)<mE$ is applied for $E<25$~meV, with the slope $m=5\times 10^{-5}$ meV$^{-2}$. (c,d) Examples of constrained (c) and unconstrained (d) fits where $S_\mathrm{HR}$ is fixed at 1.0 and 2.4, respectively.  Although lineshapes up to only 5 phonons are shown, orders up to 20 phonons are included in the model. Lineshapes are plotted as a function of the change in lattice energy during optical relaxation (i.e., $\Delta E = E_\mathrm{ZPL}-E$, where $E$ is the observed photon energy).}
  \label{PSBconstraints}
\end{figure*}}

\def\PSBfits{\begin{figure*}[htp]
%\fbox{
 \includegraphics[angle=0,scale=1,trim=0in 0in 0in 0in,clip=true]{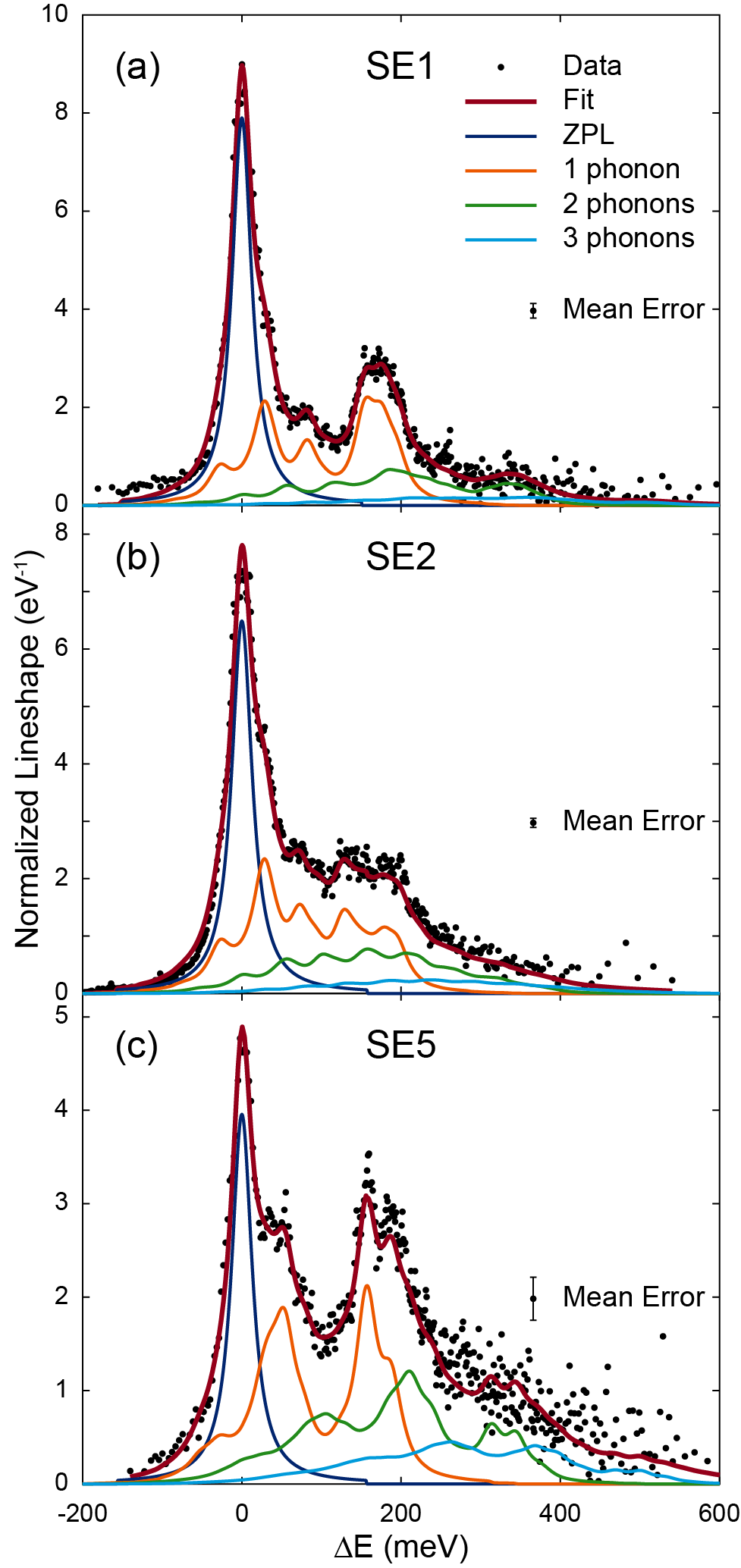}
%}
\caption{\textbf{Emission lineshapes for single emitters.} (a) SE1 (as shown in the main text), (b) SE2, and (c) SE5 (points), as a function of the change in lattice energy during optical relaxation (i.e., $\Delta E = E_\mathrm{ZPL}-E$, where $E$ is the observed photon energy). The data are binned to produce approximately uniform uncertainty, as indicated by the representative error bar. The result of a fit is shown as a thick solid line, along with the ZPL and PSB components as indicated in the legend.}
  \label{PSBfits}
\end{figure*}}
%____________________________________________________

\section{\textbf{Experimental Layout}}
\experiment

Fig. \ref{experiment} shows the general layout of the scanning confocal fluorescence microscope.  We use a 0.9 NA objective (Olympus) and 532 nm continuous (CW) excitation with $\sim 150$ $\mu$W at the sample.  The optics shown are used for all PL imaging, autocorrelation, and PL spectra, with the exception of the polarizer in the collection line, which is only in place for emission polarization dependence measurements.  Two single-photon counters are used for PL imaging and autocorrelation measurements:  Excelitas (SPCM-AQRH-14-FC) and MPD (PDM-R) detectors.  Photon autocorrelation measurements are performed using a Hanbury Brown-Twiss setup with a PicoQuant HydraHarp time correlated single-photon counting module.  PL spectra are obtained using a Princeton Instruments IsoPlane 160 spectrometer and PIXIS 100 CCD with a spectral resolution of 0.7 nm and are corrected to account for the wavelength-dependent collection efficiency of the confocal fluorescence setup.

Excitation polarization dependence is measured by rotating the linear polarization of the excitation laser using a half waveplate.  The PL is not polarization-selected; all emitted PL ($\lambda_{PL} > 550$ nm), regardless of polarization, is collected.  For the emission polarization measurements, the excitation polarization is fixed (typically at the angle which maximizes the collected PL for the defect in question) and a linear polarizer is placed in front of the detector and rotated to the desired emission polarization angles.  Polarized emission with wavelengths between 600 nm and 650 nm is measured.  Both the excitation and collection lines are corrected to account for the birefringence of the dichroic mirror and other optics in the microscope.

\section{\textbf{Sample Processing and Defect Creation}}
\materials
single-crystal h-BN purchased from HQ Graphene was exfoliated \cite{HuangSutterShiEtAl2015} onto patterned 90 nm-thick thermal SiO$_2$ on Si (Fig. \ref{materials}).  The SiO$_2$ thickness has been chosen to optimize optical contrast when imaging h-BN flakes with white light \cite{SMLL:SMLL201001628}.  To create suspended regions of h-BN, the SiO$_2$/Si support substrates have been patterned using contact photolithography and etched $\sim 5\mu$m using a fluorine-based reactive ion etch.  Patterned substrates have three distinct regions:  etched holes, free-standing SiO$_2$ (caused by under-cutting during the dry etching process), and SiO$_2$/Si regions, as denoted in the SEM image in Fig. \ref{materials}b. 

Following exfoliation, samples are annealed in Ar at 850$^{\circ}$ for 30 minutes and then undergo an O$_2$ plasma clean.  Electron bombardment is accomplished \emph{via} scanning electron microscope (SEM) imaging at 3 keV with an FEI Strata DB235 FIB SEM, after which the samples are again annealed in Ar at 850$^{\circ}$ for 30 minutes.  

\section{\textbf{Sample Characterization}}
\Raman
Fig. \ref{Raman} shows the Raman spectrum from the suspended h-BN flake shown in Fig.~1.  The peak is at 1365 cm$^{-1}$, indicative of the bulk limit.  

\AFM
Fig. \ref{AFM} shows an atomic force microscopy (AFM) image of the exfoliated h-BN flake for which single-photon emission is observed.  AFM is performed on a supported part of the exfoliated flake (dotted box in Fig. \ref{AFM}a and inset white light optical image) and measures 150 nm in height (Fig. \ref{AFM}b).   

\EBSD
Electron backscatter diffraction (EBSD) is used to verify single-crystallinity and the lattice orientation of the exfoliated h-BN flake and is performed with an FEI Quanta 600 environmental SEM at 15 keV.  The EBSD map combines two measurement metrics:  image quality, a measure of the signal to noise of the diffraction pattern (shown in greyscale), and orientation (denoted by color), whose definitions are shown in the inverse pole figure at the top right of Fig. \ref{EBSD}a.  The EBSD map in Fig. \ref{EBSD}a confirms that the exfoliated h-BN flake is oriented with the hexagonal plane parallel to the substrate.  Fits to the diffraction patterns at each point (Fig. \ref{EBSD}b) determine the in-plane crystallographic orientation.  

Fig. \ref{EBSD}c shows the $1\bar{1}00$ pole figure for the map in Fig. \ref{EBSD}a, where dense concentrations of points denote the orientation of $\langle 1\bar{1}00\rangle$ (normal to the $\{1\bar{1}00\}$ plane and $\bf{a_3}$) with the sample orientation ($\bf{A1}$ and $\bf{A2}$).  Because of the $AA'$ stacking in h-BN, we expect $60^{\circ}$ symmetry, which is reflected in the $60^{\circ}$ difference between groupings of points in Fig. \ref{EBSD}c.  In-plane orientation of the h-BN was confirmed to be constant across the sample by verifying the in-plane orientation at multiple random (high image quality) positions across the sample.  Points which do not fall into the clusters on the edge of the pole figure in Fig. \ref{EBSD}c correspond to data taken off the h-BN or over the suspended regions (where the image quality is low).  These measurements show that the h-BN flake is single-crystalline over the full supported region, and we infer that the same holds for the suspended regions as well, even though the image quality is too poor over the suspended regions to obtain a good fit to the diffraction pattern.  The h-BN lattice vector orientation shown on the EBSD map in Fig. \ref{EBSD}a reflects the orientation of the in-plane hexagonal lattice as determined from Fig. \ref{EBSD}c.

\section{\textbf{Substrate Interactions}}
\sample
Fig. \ref{sample} shows brightness differences in the PL intensity from a separate h-BN flake. Regions are visible in the images where the flake is (i) supported by the Si/SiO$_2$ substrate, (ii) supported by free-standing SiO$_2$, (iii) suspended and partially covering a hole, and (iv) fully suspended over an etched hole.  As in Fig.~1 of the main text, the brightest PL can be seen from h-BN on free-standing SiO$_2$ on the edge of an etched hole when the h-BN fully covers the hole.  Interestingly, h-BN which only partially covers a hole does not show the bright PL behavior over free-standing SiO$_2$, as evidenced by the top left hole as well as the ``U'' and the ``7''.  Furthermore, h-BN which fully suspends a hole appears to emit no PL, while h-BN which is only partially suspended or supported does show some emission.  Note that the free-standing SiO$_2$ around the small hole in the center of the region shown in Fig. \ref{sample} appears to intersect a fold or crack in the h-BN, as can be seen in the optical image in Fig. \ref{sample}b.  No PL is observed from the hole itself, over which h-BN fully suspends, but we do not observe the brightening on the free-standing SiO$_2$ which we see elsewhere, perhaps as a result of the fold or crack.  

The origin of these substrate-dependent brightness effects is not clear, but the differences suggest that interactions between the substrate and h-BN play a role in both formation and emission of visible PL.  The thin thermal oxide layer could cause constructive/destructive interference depending on the layering and hence could influence the efficiency of absorption or emission \cite{BuscemaSteeleZantEtAl2014}.  The substrate might also act as a source of impurities during annealing or irradiation, e.g., due to outgassing of oxygen by SiO$_2$. Additionally, because the etched holes only penetrate $\sim 5$ $\mu$m into the underlying Si substrate, h-BN which fully covers holes may trap air which could supply both impurities and stress during annealing, compared to partially covered holes from which the air can escape. 

\supportedsuspended
We consistently observe brighter emitters from supported h-BN when compared with suspended regions, as shown in Fig. \ref{supportedsuspended}, where emission from individual spots on supported h-BN (Si/SiO$_2$) is much brighter than from the neighboring suspended regions.  Both supported and suspended regions show many strongly excitation polarization dependent features, as evidenced by the many different colors in the polarization-resolved PL image (Fig. \ref{supportedsuspended}b) and both regions appear to have emitters which span a wide range of brightnesses (as is confirmed on a suspended region in Fig.~3c in the main text).  On average however, the contrast in brightness between supported and suspended h-BN is clear.  This observation of brighter PL from supported h-BN is somewhat counter-intuitive, as one might reasonably expect the presence of the substrate to quench PL, not enhance it.  Further studies as to emission characteristics on different types of substrates as well as the role of annealing and electron beam exposure are necessary to clarify these issues.

\section{\textbf{SE5 (A Fifth Single Emitter)}}
\H3
In addition to the four single emitters SE1--SE4 reported in Fig.~2 of the main text, we identified a fifth single emitter in the same suspended membrane.  Observations of this defect are shown in Fig. \ref{H3}.  This emitter, SE5, is excluded from the main text because its identification is somewhat ambiguous.  In the PL image of the defect (Fig. \ref{H3}a), two nearly overlapping spots are visible, one of which blinks over the course of the PL image collection. The PL spectrum in Fig. \ref{H3}b corresponds to only one of the spots; a different spectral profile is visible when both spots are present.  Identification of which spot is responsible for the autocorrelation data and PL spectrum in Fig. \ref{H3}b-c is not possible as we cannot simultaneously collect spectra and autocorrelation data.  

The excitation and emission polarization data in Fig. \ref{H3}d corresponds to the non-blinking spot in Fig. \ref{H3}a.  Note that while the emission data in Fig. \ref{H3}d fits the expected $I_{dip}(\theta)$ dependence well, the excitation data in does not.  We note that the dotted line on the excitation polarization data is not a fit; it is simply a guide to the eye.  In fact, this emitter displays remarkably similar excitation polarization dependence to SE2.  In both cases, the excitation polarization does not behave according to the expected angular dependence, but displays a clear minimum that is nearly orthogonal to that of the emission.

\section{\textbf{Emitter Stability}}
All single emitters presented here (SE1-5) were stable in ambient conditions over the course of several months.  Some emitters (SE3 and SE4) showed no evidence of blinking, whereas others (SE1, SE2, SE3, and SE5), were observed to blink on timescales from seconds to several minutes, though they spent the majority of the time in the ``on'' state.  The same is true for most spots in the suspended membrane. Whereas some spots exhibited pronounced blinking (e.g., the bright central spot in Fig.~4a of the main text), most were absolutely stable with reproducible characteristics under the measurement conditions used here.

The application of a picosecond-pulsed laser at the same average power and wavelength as the CW laser used for characterization caused a fundamental change in the emitter characteristics.  PL spectra blueshifted and emitters were significantly more likely to be in the ``off'' state when blinking. Several emitters disappeared permanently after exposure with the pulsed laser.

\section{\textbf{Autocorrelation Analysis}}
The second order autocorrelation function is given by
\begin{equation}
g^2_{ideal}(\tau) = \frac{\braket{I(t)I(t+\tau)}}{ \braket{I(t)}^2}.
\end{equation}
A single-photon source is characterized by the limiting cases $g^2(0)=0$ (only one detector can observe a single-photon at the same time) and $g^2(\infty)=1$ (uncorrelated Poissonian counts).  However, the presence of a Poissonian background (e.g., laser scatter or diffuse PL) results in a measured value of $g^2(0)> 0$  even for a single emitter.  Assuming a steady-state count rate, $S$, from the emitter on top of a Poissonian background, $B$, the measured autocorrelation function is \cite{Brouri2000}
\begin{equation}
g^2_{measured}(t) = 1-\rho^2 + \rho^2g^2_{ideal}(t)
\label{norm_limits}
\end{equation}
and the limiting cases are $g^2(0)=1-\rho^2$ and $g^2(\infty)=1$.  Note that as defined, this expression does not take into account the instrument response of the single-photon counters, which can also increase the observed value of $g^{(2)}(0)$.  

For a two-level system, the ideal autocorrelation function is
\begin{equation}
g^2_{ideal}(t) = 1-e^{-\frac{|t|}{\tau}},
\end{equation}
where $\tau = 1/(\Gamma + W_p)$, $\Gamma$ is the spontaneous decay rate, and $W_p$ is the effective pump rate. \cite{MichlerImamogluMasonEtAl2000}  In the limit that $\Gamma \gg W_p$, $\tau$ corresponds to the excited-state lifetime.  The autocorrelation measurements taken for the single emitters presented here are performed at the same power.  However, because $\tau$ is pump-power-dependent, the decay times determined through fits to the autocorrelation data may not correspond exactly to the excited-state lifetimes.  Two-level fits to measured autocorrelation data are performed using the following fit function:
\begin{equation}
g^2_{measured}(t) = A\Big(1-\rho^2e^{-\frac{|t-t_0|}{\tau}}\Big),
\end{equation}
where $t_0$ is the zero delay offset and $A$ is the uncorrelated amplitude, i.e. the steady-state number of counts as $t\rightarrow \infty$.  Normalizing the data by $A$ gives the same limiting cases as in Equation \ref{norm_limits}.

Data for which bunching is observed is fit using a three-level model:
\begin{equation}
g^2_{measured}(t) = A\Big{\{}1-\rho^2\Big[(1+a)e^{-\frac{|t-t_0|}{\tau_1}}-ae^{-\frac{|t-t_0|}{\tau_2}}\Big]\Big{\}},
\label{3level}
\end{equation}
where $\tau_1$ and $\tau_2$ are the short and long decay times and $a$ is the associated population weight. Again, the measured timescales depend on both intrinsic lifetimes of the participating electronic states together with the optical pumping rate.

The oft-quoted criterion for a single emitter in a two-level system is $g^2(0)<0.5$, but this is insufficient proof for a system exhibiting bunching (e.g., for a three-level system), where the normalized amplitude at short delays can be $>1$.  In that case, the $g^2(0)$ threshold for a single emitter is half of the maximum bunched amplitude, which is given by the limit of Equation \ref{3level} as $\tau_1 \rightarrow 0$ and $t\rightarrow 0$:
\begin{equation}
g^2_{maximum \; bunching} = 1+\rho^2a.
\end{equation}
For a three-level system therefore, the generalized requirement is that $g^2(0)<\frac{1}{2}(1+\rho^2a)$. In the case of an autocorrelation measurement fit to a two-level function, $a=0$ and the expression reduces to the usual limit of 0.5. 

\section{\textbf{Statistical Analysis of Dipole Orientation}}
\pvalue
To quantify the potential alignment of the dipole orientations with the h-BN crystal lattice, we consider a model in which excitation and emission dipoles from Fig.~4d are grouped into bins corresponding to the nearest allowed lattice vector (spaced in a 30$^\circ$ period), as a function of the crystal rotation angle, $\theta$.  For each angle setting, we calculate the chi-squared fit statistic,
\begin{equation}
\chi_m^2 = \sum_i\frac{(O_i-E_i)^2}{\sigma_i^2}, \label{chi2}
\end{equation}
where $O_i$ and $E_i$ are the observed and predicted values, respectively, and $\sigma_i$ is the uncertainty in each observation, which includes standard errors from fits to the dipole equation and systematic uncertainties associated with polarization calibration.  The $p$-value is calculated from the corresponding $\chi^2$ distribution, and corresponds to the probability that the observations would be randomly drawn from a distribution described by our model. We can also perform a fit to the data, varying $\theta$ to minimize $\chi^2_m$ and returning the apparent lattice-orientation angle, assuming our hypothesis is true.  Since some of the excitation dipoles measured in Fig.~4d may not correspond to single emitters, we also consider the case where the $N_\mathrm{bad}$ worst-aligned observations (in the sense of their contribution to $\chi^2_m$) are removed from the sum in Equation \ref{chi2}.  The $p$-value in this case is calculated assuming $N-N_\mathrm{bad}-1$ degrees of freedom, where $N=25$ is the total number of observations.

Fig.~\ref{pvalue}a shows the best-fit angle, $\theta_\mathrm{Fit}$, as a function of $N_\mathrm{bad}$.  Notably, the best-fit value agrees with the independent determination of $\theta$ using EBSD, to within experimental uncertainty, for any value of $N_\mathrm{bad}$.  The corresponding $p$-value for each fit is shown in Fig.~\ref{pvalue}b.  Conventionally, a $p$-value$<0.05$ rejects the null hypothesis (in this case, lattice locking).  If no points are excluded from the fit, that criterion is clearly satisfied ($p<10^{-7}$), so we can say with near certainty that the full set of observations does not indicate universal lattice locking.  However, $p$ rapidly rises as points are removed from the fit, passing the rejection threshold for $N_\mathrm{bad}>4$.  

Taken alone, the results of the statistical analysis are only sufficient to show that the hypothesis of lattice-locking cannot be disproved by these data.  However, the fact that the analysis finds that the most likely grouping occurs at angles corresponding to the crystallographic axes determined \emph{via} EBSD is compelling evidence that at least a subset of defects in suspended h-BN have dipoles which align with the lattice vectors.  To test the strength of this finding, we have considered an alternate null model where randomly generated values are used in place of the data, with the uncertainties unchanged.  We find that the probability of random data reproducing these results, in the sense that $p>$0.05 and $\theta_\mathrm{fit}=\theta_\mathrm{EBSD}$ within uncertainties, is $\sim$20\%.  Hence, this analysis is only suggestive of dipole-lattice alignment in a limited sense, and does not prove it to high confidence.  Further measurements, particularly of individual classes of emitter, will be needed to clarify this issue further.

\section{\textbf{Finite Temperature Phonon Side Band Analysis}}
\PSBconstraints 
Measured PL spectra as a function of wavelength are corrected for the photon detection efficiency of our confocal setup (determined by observing a calibrated light source in place of the sample) and binned to obtain the spectral distribution function ($S(\lambda)$).  Next, we calculate the distributions $S(E)$ and $L(E)$, where 
\begin{equation}
S(E)=S(\lambda)(hc/E^2)
\end{equation}
accounts for $|\frac{d\lambda}{dE}|=\frac{hc}{E^2}$ and corresponds to the underlying spectral probability distribution function of emitting phonons with energy $E$ and
\begin{equation}
L(E) = S(E)/E^3
\end{equation}
is the spectral lineshape which accounts for the $E^3$ energy dependence due to spontaneous emission. Only the function $L(E)$ can be directly compared with the predictions of the Huang-Rhys theory, developed below.  

In the conversions between different spectral functions, we propagate experimental uncertainties associated with photon shot noise, background noise in the CCD detector, and long-time fluctuations observed in the spectral shape between exposures.  The combined effects of applying calibration corrections for the wavelength-dependent quantum efficiency of our setup together with the lineshape conversions tend to amplify the noise at longer wavelengths.  To compensate for this, we adjust the binning as a function of wavelength to yield approximately uniform uncertainty for each data point.

Following the theory developed by Maradudin \cite{Maradudin1966} and applied extensively to diamond defects by Davies \cite{Davies1974}, the predicted spectral lineshape as a function of lattice energy, $E$, is modeled by
\begin{equation}
L(E) = e^{-S_{HR}}I_0(E)+I_0(E)\otimes I_{PSB}(E),
\end{equation}
where $S_{HR}$ is the Huang-Rhys factor, $I_0(E)$ is the normalized ZPL lineshape, and $I_{PSB}$ is the predicted PSB when the broadening of the ZPL is neglected. The convolution with $I_0(E)$ accounts approximately for both of these effects \cite{Davies1974}.

The PSB can be divided into contributions involving integer numbers of phonons:
\begin{equation}
I_{PSB}(E) = \sum_n e^{-S_{HR}}\frac{S_{HR}^n}{n!}I_n(E),
\end{equation}
where $I_n = I_1 \otimes I_{n-1}$ are the normalized $n$-phonon probability distributions.  At finite temperatures, the 1-phonon distribution is defined in terms of an underlying distribution function describing the electron-phonon coupling for phonon energy $E$, $S(E)$,  and the Bose-Einstein phonon distribution function, $n(E,T)=1/\big(e^{E/kT}-1\big)$, in a piecewise fashion as
\begin{equation}
\left\{\begin{matrix}
I_1(E)=A[n(E,T)+1]S(E) & \mbox{for }E>0\\ 
I_1(E)=An(-E,T)S(-E) & \mbox{for }E<0,
\end{matrix}\right.
\end{equation}
where $A$ is a normalization constant.  $S(E)$ is defined for energies $E\in [0,E_{max}]$, where $E_{max}$ is typically the maximum phonon energy, which, in h-BN, is $\sim$200 meV\cite{PhysRevLett.98.095503}.  We use a Lorentzian lineshape to model the room-temperature ZPL since weak intra-layer coupling in h-BN mean that thermal broadening effects should dominate \cite{Davies1974,PhysRev.133.A1029,PhysRevLett.98.095503}.  

We apply the model to analyze the spectra for SE1, SE2, and SE5, for which the ZPL can be clearly identified. Fits have the following free parameters: $S_{HR}$, the FWHM of the Lorentzian lineshape for the ZPL ($\Gamma_{ZPL}$), the ZPL energy ($E_{ZPL}$), and the phonon spectral function $S(E)$, which is parameterized by discrete values at energies $\{\frac{dE}{2},\frac{3dE}{2},\ldots,E_\mathrm{max}-\frac{dE}{2}\}$, for some energy resolution, $dE$. Fits in this work are performed with $dE=5$~meV.  Note that $S(0)=0$ is required in order to yield a finite value for $I_1(0)$.  

When $S(E)$ is completely unconstrained, strong covariances exist in the model between $S_{HR}$, $\Gamma_{ZPL}$, and the low-energy components of $S(E)$ for $E\lesssim \Gamma_{ZPL}$. The effect of this covariance is highlighted in Fig.~\ref{PSBconstraints}. The chi-squared fit statistic is remarkably insensitive to large variations in the fixed value of $S_\mathrm{HR}$ (Fig.~\ref{PSBconstraints}a), indicating that reasonable fits can be obtained for nearly any value $S_\mathrm{HR}\gtrsim 1$.  For $S_\mathrm{HR}<1$, the covariance is mainly with $\Gamma_\mathrm{ZPL}$ as the fit tries unsuccessfully to match the shape of the dominant ZPL peak.  Around $S_\mathrm{HR}=1$, both constrained and unconstrained fits arrive at the traditional picture of a ZPL and spectrally resolved PSB. For $S_\mathrm{HR}>1$, the covariance shifts to $S(E)$, which develops large contributions at low energies that produce a ``ZPL-like'' peak in the PSB (Fig.~\ref{PSBconstraints}d) that adds to the suppressed ZPL.
\PSBfits 

Even though the unconstrained model faithfully reproduces the observed emission lineshape for $S_{HR}>1$, the corresponding large weight of $S(E)$ at low phonon energies is probably an artifact of the fitting and not physical. The typical expectation for coupling to low-energy acoustic phonons in 3D semiconductors is that $S(E)\propto E$.  We note that it has recently been suggested that coupling to low-energy phonons can be enhanced due to piezoelectric effects in h-BN at the 2D limit \cite{2016arXiv160504445J}.  Our experiments, however, concern a 150-nm-thick film composed of hundreds of atomic layers, well within the bulk 3D limit.  In this regime, h-BN possesses inversion symmetry and is not piezoelectric, and as a result we expect deformation potential coupling to dominate interactions with acoustic phonons at low $E$ energies, leading to a linear dependence for $S(E)$.

To eliminate this covariance between $S_\mathrm{HR}$ and $S(E)$, we apply a constraint such that $S(E)<mE$ for a chosen slope, $m$, up to a cutoff energy, $E_\mathrm{cutoff}$.  These constraints result in fits where $S_\mathrm{HR}\sim 1$, and the Lorentzian ZPL dominates the low energy behavior of $L(E)$ (Fig. \ref{PSBconstraints}c).  The constraints have little effect on the high energy shape of the PSB.  To estimate the values and confidence intervals for $S_\mathrm{HR}$ and $\Gamma_\mathrm{ZPL}$ listed in Table 1 of the main text, we performed fits for a range of settings for $m$ and $E_\mathrm{cutoff}$ that vary the impact of the constraint.  Specifically, we considered values for $m$ between $1\times 10^{-5}$ meV$^{-2}$ (highly constrained) and $2\times 10^{-4}$ meV$^{-2}$ (weakly constrained), and $E_\mathrm{cutoff}$ ranging from 20-30 meV, which is below the large density of states of acoustic phonons at the $K$-point in h-BN\cite{PhysRevLett.98.095503}.  Fig.~\ref{PSBfits} shows the full results of fits to SE1, SE2, and SE5, whose 1-phonon lineshape function is plotted in Fig.~3 of the main text.  These fits use $m=5\times 10^{-5}$ meV$^{-2}$ and $E_\mathrm{cutoff}=25$~meV.

%\bibliography{Exarhos_supplemental_bibliography}